# Recent Advances in Unconventional Ferroelectrics and Multiferroics

*Hongyu Yu, Junyi Ji, Wei Luo\*, Xingao Gong, Hongjun Xiang\**

Hongyu Yu, Xingao Gong, Hongjun Xiang

Key Laboratory of Computational Physical Sciences (Ministry of Education), Institute of Computational Physical Sciences, State Key Laboratory of Surface Physics, and Department of Physics, Fudan University, Shanghai 200433, China

E-mail: hxiang@fudan.edu.cn

Junyi Ji

Beijing National Laboratory for Condensed Matter Physics and Institute of Physics, Chinese Academy of Sciences, Beijing 100190, China

Wei Luo

Smart Ferroic Materials Center, Physics Department and Institute for Nanoscience and Engineering, University of Arknasas Fayetteville, Arkansas 72701, USA

E-mail: weil@uark.edu

Funding: The National Key R&D Program of China (No. 2022YFA1402901), NSFC (Grant Nos. 12188101), Shanghai Science and Technology Program (No. 23JC1400900), the Guangdong Major Project of the Basic and Applied Basic Research (Future functional materials under extreme conditions–2021B0301030005), Shanghai Pilot Program for Basic Research—Fudan University 21TQ1400100 (23TQ017), China National Postdoctoral Program for Innovative Talents (BX20230408), the robotic AI-Scientist platform of Chinese Academy of Science, and New Cornerstone Science Foundation.

**Abstract:** Emerging ferroic materials may pave a new way to next-generation nanoelectronic and spintronic devices due to their interesting physical properties. Here, we systematically review unconventional ferroelectric systems, from Hf-based and elementary ferroelectrics to stacking ferroelectricity, polar metallicity, fractional quantum ferroelectricity, wurtzite-type ferroelectricity, and freestanding membranes ferroelectricity. Moreover, multiferroic materials are reviewed, particularly the interplay between novel magnetic states and ferroelectricity, as well as ferrovalley-ferroelectric coupling. Finally, we conclude by discussing current challenges and future opportunities in this field.

# 1. Introduction

Ferroic materials, characterized by their spontaneous ordering of electric dipoles, magnetization, or strain, have long been at the forefront of technological innovation, underpinning a vast array of applications from data storage to sensors and actuators[1–12]. Traditional ferroelectrics, ferromagnets, and ferroelastics have been extensively studied and implemented. However, recent years have witnessed a remarkable expansion in the realm of ferroics, fueled by the discovery of novel phenomena and materials that challenge conventional understanding[13–18]. The exploration of novel ferroic materials is not merely about discovering new compounds; it is fundamentally about understanding and manipulating the intricate interplay of electronic, magnetic, and structural degrees of freedom at the nanoscale. This burgeoning of novel ferroic materials holds immense promise for advancing next-generation nanoelectronic and spintronic technologies, offering pathways to overcome limitations in current device paradigms[19].

This review aims to provide a comprehensive overview of the recent progress in the exciting field of novel ferroic materials. We will systematically explore several classes of unconventional ferroelectric systems that have emerged as focal points of contemporary research. These include Hf-based ferroelectrics, which have garnered significant attention due to their compatibility with silicon technology[20], and elementary ferroelectrics that challenge the traditional understanding of ferroelectricity in simple materials[21–23,15]. We will also discuss stacking ferroelectricity, a fascinating phenomenon arising from interlayer interactions in van der Waals heterostructures[24–29], polar metallicity, which opens up new avenues for integrating ferroicity with conductivity[3,30], fractional quantum ferroelectricity, representing a conceptually novel state of matter[17,31], wurtzite-based ferroelectrics which are characterized by their tetrahedral coordination[32–34], and the unique physics in freestanding ferroelectric membranes[35–38]. Beyond ferroelectricity, the review will extend its scope to multiferroic materials, where the simultaneous presence of two or more ferroic orders leads to intriguing coupling effects and functionalities[39]. We will particularly focus on the interplay between novel magnetic states, such as magnetic skyrmions[40] and altermagnetism[14,41–43], and ferroelectricity, highlighting the potential for manipulating ferroelectric order through magnetic stimuli and vice versa. Furthermore, we will explore the emerging field of ferrovalley-ferroelectric coupling, which leverages the valley degree of freedom to achieve novel functionalities[16,44–48]. Finally, we will conclude by discussing the key challenges and future opportunities that lie ahead. The review is structured as follows: Section 2 will be dedicated to unconventional ferroelectrics, while Section 3 will focus on multiferroic systems with the coupling between different ferroic orders and electronic/magnetic/valley degrees of freedom. At last, Section 4 will summarize the key findings and discuss future perspectives in this rapidly advancing domain.

## 2. Unconventional Ferroelectrics

The field of ferroelectrics has traditionally been dominated by perovskite oxides and related compounds. Conventional ferroelectric materials, such as perovskite oxides like $BaTiO_3$ (B-site driven) and $BiFeO_3$ (A-site lone-pair driven), as well as geometric ferroelectrics (e.g., hexagonal manganites), spin-order induced (Type-II multiferroics), and charge-order driven ferroelectrics, have been extensively studied[49]. These materials, while diverse in their specific polarization mechanisms, generally involve ionic displacements leading to a switchable spontaneous polarization. Beyond these established classes, recent years have seen an explosion of research into "unconventional" systems that challenge traditional definitions and expand the landscape of polar materials. A material is termed "polar" if it possesses a spontaneous electric dipole moment due to a lack of inversion symmetry in its crystal structure. However, to be classified as "ferroelectric", a polar material must exhibit a spontaneous polarization that can be reoriented or switched between two or more stable states by an externally applied electric field. This switchability is the hallmark of ferroelectricity and is essential for most memory and logic applications. A diverse landscape of unconventional ferroelectric systems has emerged shown in Figure 1, pushing the boundaries of our understanding and offering exciting new avenues for materials innovation. Researchers have explored several classes of unconventional ferroelectrics. These unconventional ferroelectrics deviate from the traditional paradigms in various aspects, including their composition, polarization mechanisms, and functional properties. The exploration of these novel systems is driven by the quest for enhanced performance characteristics, such as higher Curie temperatures, larger polarizations, and improved switching speeds, as well as the desire to uncover fundamentally new functionalities and broaden the application scope of ferroelectric materials. Meanwhile, it should be noted that in some unconventional ferroelectrics, switchable ferroelectricity remains to be unambiguously demonstrated. The following subsections will delve into several prominent classes of unconventional ferroelectrics, starting with the technologically significant field of Hf-based ferroelectricity, and subsequently examining elementary ferroelectrics, stacking ferroelectricity, polar metallicity, fractional quantum ferroelectricity, wurtzite ferroelectricity, and freestanding ferroelectric membranes, each representing a distinct and fascinating facet of this rapidly expanding area of research.

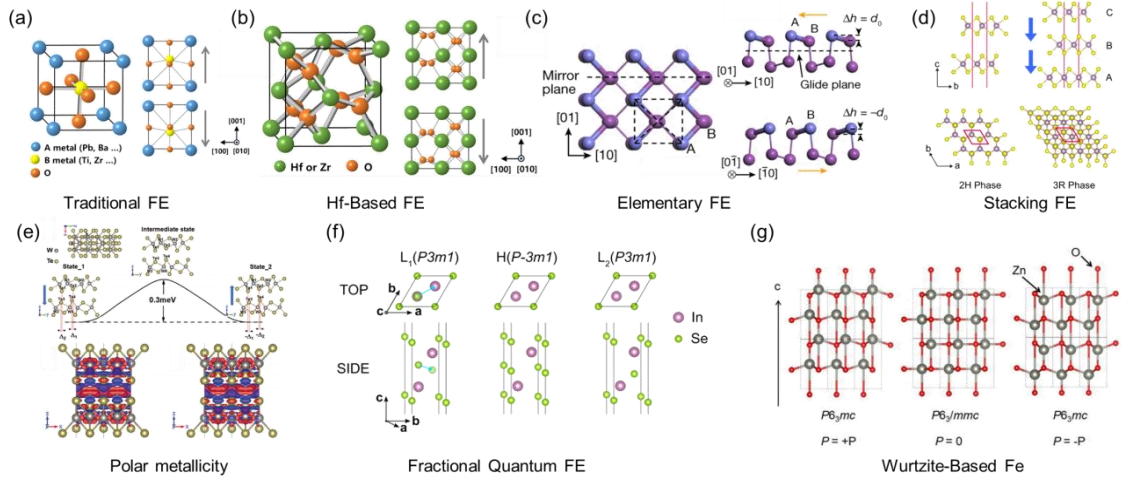

**Figure 1.** The crystal structures and ferroelectric mechanisms of various ferroelectric materials. (a) The perovskite cubic $ABO_3$ crystal structure with traditional FE is presented. The figure on the right is a schematic diagram depicting the structures of two opposite polarization states of perovskite ferroelectrics. Their polarization states are contingent upon the offset position of the B metal ion at the body center. Reproduced with permission.[20] Copyright 2025, Elsevier. (b) The figure on the left is a crystal structure diagram of the orthorhombic phase of $HfO_2$. The figure on the right is a schematic diagram of two opposite polarization states. The polarization state is dependent on the offset position of the four oxygen ions within the unit cell. Reproduced with permission.[20] Copyright 2025, Elsevier. (c) The lattice structure of single-layer black phosphorous-like-Bi with elementary FE. The uppermost Bi atoms are colored light blue. Reproduced with permission.[15] Copyright 2023, Springer Nature Limited. (d) Top and side view schematics of the atomic structures of 2H and 3R $MoS_2$ with stacking FE are presented. The yellow and purple spheres denote the sulfur and molybdenum atoms, respectively. The unit cells are demarcated by red diamond-shaped boxes. The blue arrows signify the direction of spontaneous polarization. Reproduced with permission.[50] Copyright 2022, Springer Nature Limited. (e) The ferroelectric switching pathway of bilayer $WTe_2$ with polar metallicity. Reproduced with permission.[51] Copyright 2019, Royal Society of Chemistry. (f) Top and side views of monolayer α-$In_2Se_3$ featuring fractional quantum ferroelectricity (FE). The in-plane displacement of Se during the ferroelectric phase transition is indicated by the blue arrow. Reproduced with permission.[17] Copyright 2024, Springer Nature Limited. (g) Schematic illustration of polarity inversion in a wurtzite FE via an intermediate nonpolar hexagonal structure. Reproduced with permission.[33] Copyright 2014 AIP Publishing LLC.

## 2.1. Hf-based ferroelectricity

Hafnium oxide (HfO$_2$)-based materials, particularly Hf$_x$Zr$_{1-x}$O$_2$, have emerged as important contenders in the realm of next-generation electronics due to their unique combination of robust ferroelectric properties and exceptional compatibility with complementary metal-oxide-semiconductor (CMOS) technology[52–56,20]. This convergence of ferroelectricity with CMOS-compatible materials has ignited significant interest across academia and industry, positioning Hf-based ferroelectrics as pivotal in the advancement of nanoelectronics and neuromorphic computing[13,57–59]. These materials present a compelling pathway to address the growing demands for high-performance, energy-efficient electronic devices, offering a unique blend of properties that promise to overcome limitations inherent in conventional technologies.

The fundamental nature of ferroelectricity in HfO$_2$ distinguishes it from traditional ferroelectric mechanisms. Departing from the conventional understanding where cation displacements are the primary drivers of ferroelectricity, HfO$_2$ exhibits an unconventional mechanism centered on the anionic sublattice. As shown in Figure 2, it is the oxygen ions, rather than the hafnium ions, that undergo significant displacement within the crystal lattice, leading to the emergence of ferroelectric behavior[13]. This anionic displacement mechanism sets Hf-based ferroelectrics apart and underpins their unique characteristics, prompting intense scientific inquiry into the intricacies of their behavior and potential. This mechanism implies that the ferroelectric phase in HfO$_2$ is not thermodynamically stable in its bulk form, unlike classical ferroelectrics, which naturally exist in a ferroelectric state at ambient conditions[59,60]. Instead, the stabilization of ferroelectricity in HfO$_2$ is a delicate process, contingent upon a confluence of factors that carefully tune the material's energy landscape. These factors encompass surface energy considerations, where interfacial effects become dominant at nanoscale dimensions, and the introduction of chemical pressure through precise doping, which modifies the lattice structure and electronic properties of the material[61,62]. Furthermore, careful control over processing conditions, particularly annealing temperature and atmosphere, is crucial in dictating the phase formation kinetics and ensuring the emergence of the desired ferroelectric phase.

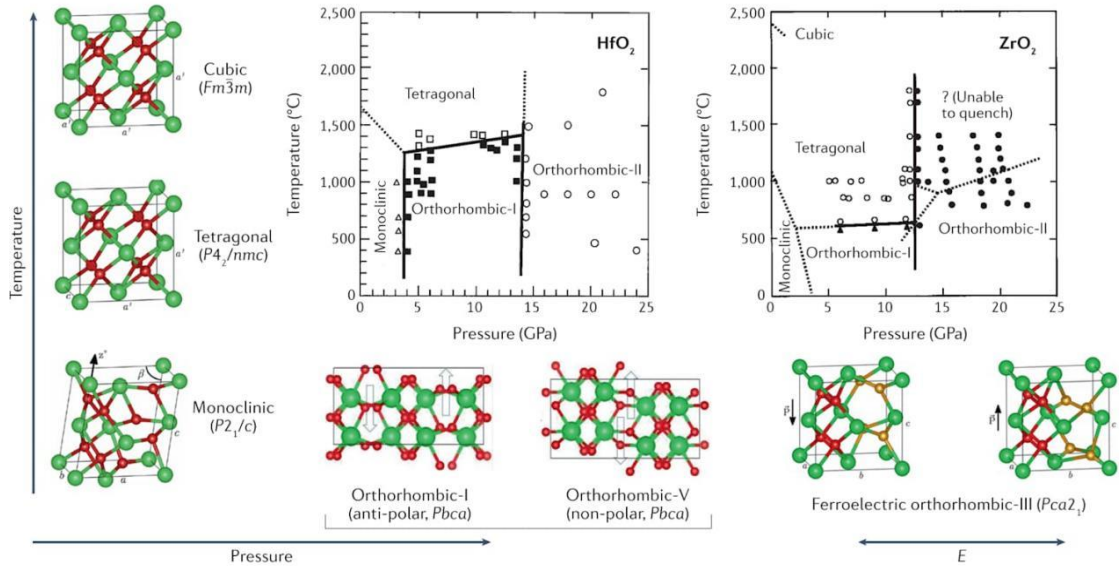

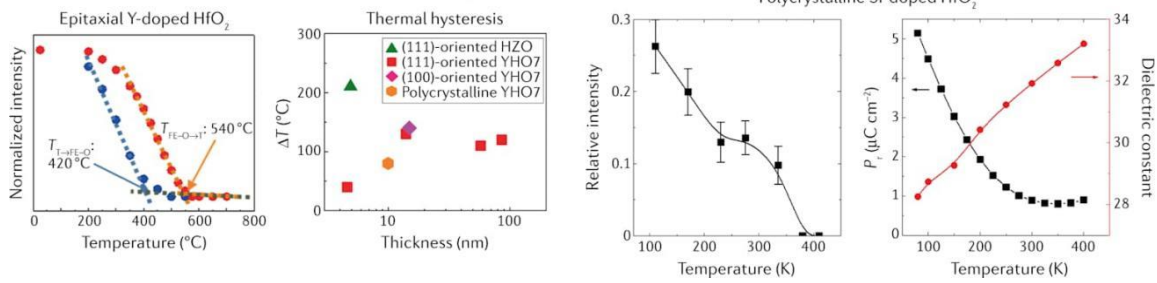

**Figure 2.** Phase diagrams and structural transitions in hafnium oxide ferroelectrics. (a) Temperature-pressure phase diagrams of $HfO_2$ and $ZrO_2$ showing various crystal structures, including cubic, tetragonal, monoclinic, and ferroelectric orthorhombic-III phases. (b) Phase transition characteristics from the ferroelectric orthorhombic-III phase in epitaxial Y-doped $HfO_2$ and polycrystalline Si-doped $HfO_2$, demonstrating thermal hysteresis behavior. Reproduced with permission. Reproduced with permission.[63] Copyright 2022, Springer Nature Limited.

Several intrinsic properties render Hf-based ferroelectrics particularly attractive for advanced technological applications. One such property is their high coercive field ($E_c$), which is typically an order of magnitude larger than that of traditional perovskite ferroelectrics[64]. The high coercive field, while requiring higher operating voltages, actually improves data retention and stability as the key features for non-volatile memory devices. Additionally, $HfO_2$'s wide bandgap of over 5 eV helps reduce leakage currents, a common problem in nanoscale ferroelectrics. This intrinsic low leakage characteristic is instrumental in enabling the fabrication of smaller, more reliable devices and facilitates low-power operation, aligning with the stringent energy efficiency demands of modern electronics[65]. One of the most important intrinsic properties of Hf-based ferroelectrics is their exceptional thickness scalability. Unlike conventional ferroelectrics that lose their ferroelectric properties below a critical thickness, Hf-based materials maintain robust ferroelectricity even when scaled down to the nanometer regime[66]. This remarkable scalability is a decisive advantage, enabling the development of high-density memory architectures and facilitating seamless integration into advanced semiconductor processes. Furthermore, the excellent compatibility of Hf-based ferroelectrics with CMOS (Complementary Metal-Oxide-Semiconductor) technology represents a watershed moment in the field[57,55,65]. This compatibility allows for their direct integration into existing silicon-based microelectronics fabrication lines, obviating the need for specialized processing techniques and significantly reducing manufacturing costs, thereby accelerating the translation of Hf-based ferroelectrics from laboratory research to industrial applications[67].

Phase stabilization in $HfO_2$, given the metastability of its ferroelectric phase, is a critical aspect of materials engineering[58]. Doping stands out as a primary method for achieving this stabilization[61]. By introducing dopants such as zirconium (Zr)[68], silicon (Si)[69], aluminum (Al)[70], lanthanum (La)[71], and yttrium (Y)[72] into the $HfO_2$ lattice, chemical pressure is exerted, and the energy landscape of the material is modified, effectively lowering the energy barrier for the formation of the ferroelectric orthorhombic phase relative to the thermodynamically stable monoclinic phase[61]. The selection of dopant and the careful calibration of its concentration are paramount, as they directly influence the resulting ferroelectric properties and the overall phase stability of the material. Beyond doping, the meticulous engineering of oxygen vacancy concentrations emerges as another crucial factor[58,73,74]. While oxygen vacancies are often viewed as detrimental defects in traditional ferroelectrics, in $HfO_2$, they exhibit a more nuanced behavior. Under precisely controlled conditions, oxygen vacancies can contribute to the stabilization of the ferroelectric phase[58]. Techniques such as oxygen scavenging, which involves using materials with a high oxygen affinity to control oxygen stoichiometry during annealing, and the precise regulation of oxygen partial pressure during deposition are employed to fine-tune oxygen vacancy concentrations and optimize ferroelectric performance. Mechanical stress and strain, whether inherent to the film growth process or externally applied through substrates or electrodes, exert a profound influence on the phase stability of $HfO_2$[75]. Tensile stress, for example, has been shown to promote the formation of the ferroelectric orthorhombic phase, whereas compressive stress may favor alternative phases[76,77]. The careful management of stress through judicious substrate selection, electrode engineering, and optimization of film deposition techniques is therefore indispensable for achieving targeted phase stabilization. Finally, the precise control of thermal annealing processes is of paramount importance[67,68,70,73,78]. Tailoring annealing temperature, duration, and atmosphere allows for the meticulous manipulation of crystallization kinetics and phase formation in $HfO_2$ films. Rapid thermal annealing and low-temperature annealing techniques are frequently employed to ensure the formation of the desired ferroelectric phase while concurrently minimizing undesirable phase transformations and the generation of defects that could compromise device performance[55]. While the polar orthorhombic phase is widely considered the primary contributor to ferroelectricity in $HfO_2$-based films, other polar phases, such as rhombohedral R3m or R3 structures, have also been proposed or observed, particularly under specific strain conditions or in certain doped compositions[79]. However, recent first-principles studies suggest that these rhombohedral phases in $HfO_2$ may have poorer phase stability or ferroelectric cycling endurance compared to the orthorhombic phase[80], potentially decaying to lower-energy tetragonal or monoclinic phases , making the orthorhombic phase the more robust candidate for practical applications.

Domain walls, the interfaces separating regions of differing polarization orientation within $HfO_2$ ferroelectrics, exhibit complex dynamics that significantly impact device behavior, particularly concerning fatigue and imprint phenomena[64,81,82]. In contrast to conventional ferroelectrics, domain walls in $HfO_2$ are characterized by a higher energy barrier for propagation, implying a distinct and less facile switching mechanism. Through advanced characterization techniques like piezoresponse force microscopy (PFM), coupled with sophisticated theoretical modeling, researchers have gained insights into the intricate switching pathways within $HfO_2$, revealing the dominance of E-path and T-path switching mechanisms[83–85]. The E-path mechanism involves the direct switching between polar and nonpolar half-cells within the $HfO_2$ lattice, whereas the T-path mechanism is associated with domain wall motion and the nucleation of new domains. A comprehensive understanding of these domain dynamics[81] and switching mechanisms is crucial for devising effective strategies to mitigate reliability concerns and optimize the switching characteristics of $HfO_2$-based devices, paving the way for their widespread adoption in practical applications.

The confluence of CMOS compatibility, scalability, and unique ferroelectric properties positions Hf-based ferroelectrics as enabling materials for a diverse spectrum of advanced applications. In the realm of non-volatile memory, Ferroelectric RAM (FeRAM) based on $HfO_2$ emerges as a compelling alternative to traditional flash memory, offering faster access times, reduced power consumption, and enhanced endurance, thereby addressing critical performance bottlenecks in memory technology[55]. Furthermore, the analog switching characteristics and multi-level conductance states inherent to Hf-based ferroelectric devices render them exceptionally well-suited for neuromorphic computing architectures[57]. These devices hold the key to emulating the intricate synaptic plasticity of the human brain, enabling the development of energy-efficient neuromorphic hardware for advanced artificial intelligence applications.

**2.2. Elementary ferroelectricity**

For many years, the prevailing understanding of ferroelectricity dictated that it was a phenomenon exclusive to compound materials, requiring the presence of at least two different elements. This was thought necessary to create an inherent asymmetry within the material, leading to spontaneous polarization. This asymmetry, typically arising from differences in electronegativity between constituent atoms, was considered crucial for the formation of electric dipoles and their subsequent alignment into a ferroelectric state. However, recent theoretical predictions and experimental confirmations have dramatically altered this perspective, demonstrating the existence of intrinsic ferroelectricity in materials composed of a single element[15,86,21,22,87–90,23,91]. This discovery not only expands our understanding of ferroic materials but also opens new avenues for research and potential applications, leveraging the simplicity and unique properties of these materials.

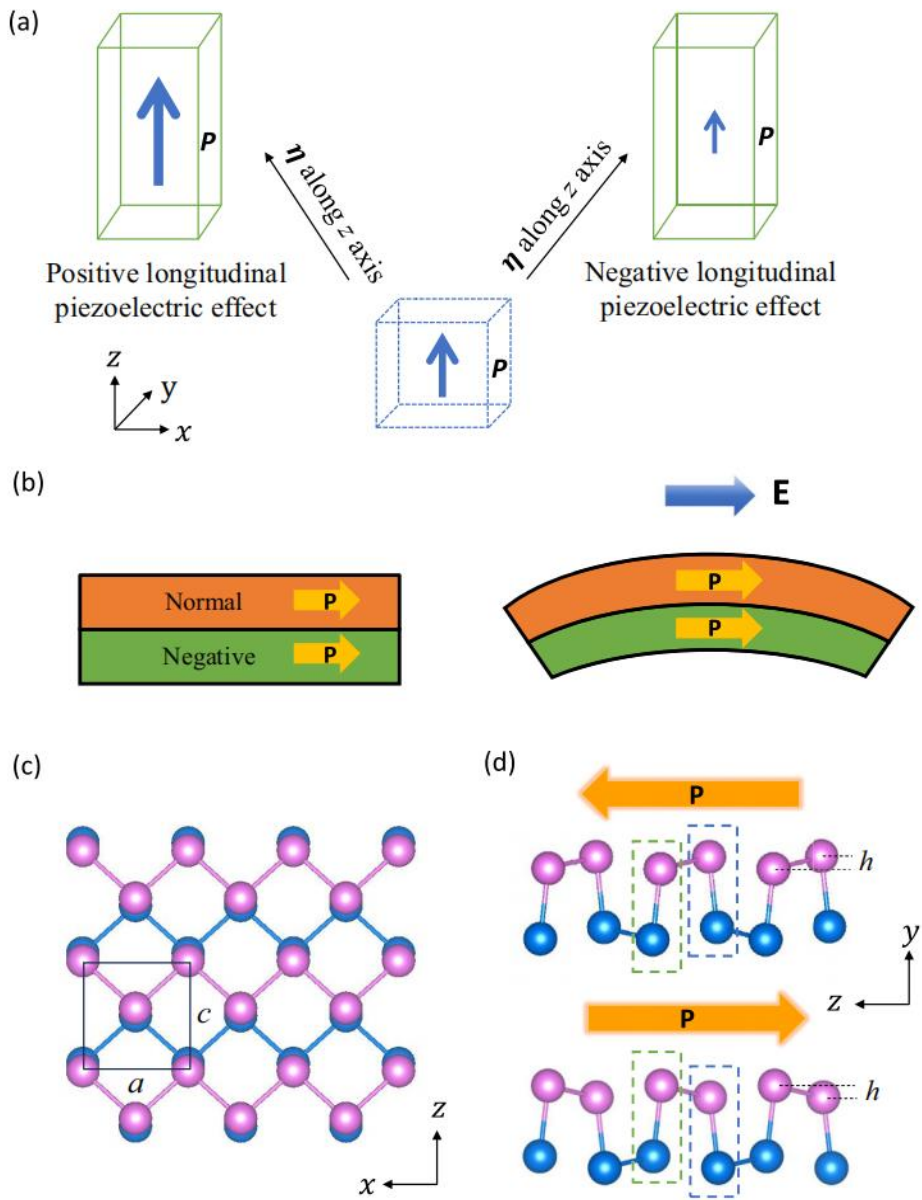

**Figure 3.** (a) Schematic of positive and negative longitudinal piezoelectric effects. The assumed applied strain is along the z-axis. (b) Schematic of the mechanical bending in a normal/negative piezoelectric heterostructure. (c) and (d) Structure of the ferroelectric α-Bi monolayer. The upper and lower Bi atoms are distinguished by colors. (c) Top view. The primitive cell is indicated by the black rectangle. (d) Side views of the two degenerate polar states. The buckling h corresponds to the intercolumn (dashed boxes) sliding. Such sliding can induce an in-plane polarization, similar to the sliding ferroelectric mechanism in vdW layered structures. For piezoelectric tensor analysis, the in-plane polar axis is defined as the z-axis. Reproduced with permission.[92] Copyright 2023, American Physical Society.

Pioneering studies predicted ferroelectricity in two-dimensional (2D) monolayers of Group V elements[22] [89] [15], including arsenic (As), antimony (Sb), and bismuth (Bi). These studies highlighted the importance of spontaneous lattice distortion, such as atomic layer buckling, and the resulting breaking of centrosymmetry, as key factors in inducing ferroelectricity in these elemental systems. The non-centrosymmetric arrangement of atoms, resulting from these distortions, creates a net electric dipole moment within the unit cell, leading to macroscopic spontaneous polarization. Importantly, the predicted Curie temperatures ($T_c$) for these materials often exceeded room temperature, suggesting their potential for practical applications. The lone-pair electrons, a characteristic feature of Group V elements, drive these distortions and contributing to the spontaneous polarization. A significant breakthrough was the experimental verification of ferroelectric switching in Bi monolayers[15]. Using Scanning probe microscopy (SPM), researchers directly visualized and manipulated ferroelectric domains in Bi monolayers, applying in-plane electric fields with a scanning tunneling microscope tip to induce reversible polarization switching. This provided unambiguous evidence for the theoretically predicted ferroelectricity in elemental Bi[22]. The observed switching, characterized by hysteresis loops and reversible domain reversals, was consistent with theoretical models, reinforcing the proposed mechanisms. Crucially, these studies confirmed the stability of the ferroelectric phase in Bi monolayers at room temperature[15], enhancing their potential for practical applications. Another significant achievement was the discovery of room-temperature ferroelectricity in quasi-one-dimensional Te nanowires[21,23]. Researchers probed the local ferroelectric properties of Te nanowires with PFM, observing clear hysteresis loops and domain reversals, confirming the presence of macroscopic spontaneous polarization[23]. This observation, combined with the high carrier mobility and resistive switching capabilities of Te nanowires[23], opened up possibilities for their use in high-density data storage and novel electronic devices. Density functional theory (DFT) calculations suggested that Te multilayers could exhibit in-plane polarization due to interlayer interactions associated with lone pair electrons, a prediction later confirmed experimentally[23]. Even silicon (Si), the cornerstone of modern electronics, was suggested with potential ferroelectricity. Computational studies predicted metastable ferroelectric phases in reconstructed silicon bilayers, demonstrating that ferroelectricity is not limited to specific elemental groups but can arise in Group IV semiconductors under specific structural conditions[90]. These studies greatly expand our understanding of this unconventional phenomenon.

The mechanisms driving ferroelectricity in elemental materials differ significantly from those in conventional compound ferroelectrics. In conventional compound ferroelectrics, polarization typically arises from the relative displacement of cations and anions, leading to a net dipole moment within the unit cell[49]. Several distinct mechanisms contribute to this. In many perovskite oxides like $BaTiO_3$, ferroelectricity is "B-site driven", where a small, highly charged cation (e.g., $Ti^{4+}$) with a $d^0$ electronic configuration displaces off-center within its oxygen octahedron. In other perovskites like $BiFeO_3$ or $PbTiO_3$, the stereochemically active $s^2$ lone pair on the A-site cation (e.g., $Bi^{3+}$, $Pb^{2+}$) is the primary driving force for the polar distortion. Geometric ferroelectricity, observed in materials like hexagonal $YMnO_3$, arises from rigid rotations and tilting of polyhedra that break inversion symmetry. Organic ferroelectrics, such as polyvinylidene fluoride (PVDF) and its copolymers, exhibit polarization due to the alignment of molecular dipoles associated with C-F and C-H bonds. Furthermore, Type-II (spin-induced) multiferroics exhibit ferroelectricity that originates from specific magnetic ordering, which can be explained by the unified polarization model[93–96]. The charge-order–driven ferroelectricity arises from electronic ordering phenomena. These diverse mechanisms in conventional ferroelectrics provide a rich context for understanding the even more varied origins of polarity in the unconventional systems discussed below. In elemental systems, the absence of dissimilar ions necessitates alternative mechanisms to break centrosymmetry. A common theme is the role of structural distortions in breaking the centrosymmetry of the crystal lattice. These distortions, such as atomic layer buckling in 2D monolayers or interlayer sliding in multilayers, create an imbalance in charge distribution, leading to a net electric dipole moment and ferroelectric order[21,90]. Lone-pair electrons play a vital role in driving these distortions and contributing to polarization. In Group V monolayers, stereochemically active lone-pair electrons are believed to stabilize the buckled structure, influencing charge distribution and contributing to spontaneous polarization[89,22]. In Te nanowires, interactions between lone-pair electrons on adjacent chains are proposed to drive the ion displacements leading to vertical polarization[23]. The observation of charged domain walls in Bi monolayers, deviating from typical behavior in compound ferroelectrics, highlights the distinct nature of the underlying mechanisms[15]. These charged domain walls are stabilized by a balance of strain and electrostatic energies, with strain energy dominating over electrostatic repulsion[91,97]. Ferroelectric switching mechanisms in elemental materials involve complex interactions between electronic structure, lattice dynamics, and applied fields. Further research, using advanced modeling and time-resolved experimental techniques, is needed to fully understand these switching processes.

As for polarization, theoretical predictions for Group V monolayers (As, Sb, Bi) suggest the value in the order of 0.02 μC/cm², with $T_c$ estimates exceeding room temperature due to strong lone-pair-driven structural distortions[22,23]. Experimental verification in Bi monolayers via SPM revealed reversible polarization switching, though quantitative polarization measurements remain challenging in ultrathin geometries. Tellurium nanowires exhibit clearer hysteresis loops via PFM, with reported polarizations of ~0.026 μC/cm² and robust room-temperature stability[23]. This value is orders of magnitude smaller than the 50-100 μC/cm² typically observed in optimized perovskite thin films. A key challenge in characterizing elemental ferroelectrics lies in disentangling intrinsic displacement currents from leakage contributions, particularly in systems with high carrier mobility. As highlighted by Scott[98], ordinary household objects can exhibit closed hysteresis loops that superficially resemble ferroelectric behavior but arise entirely from leakage currents and non-ferroelectric mechanisms. Several advanced techniques have been developed to separate intrinsic ferroelectric switching currents from leakage contributions. The dual-frequency measurement approach applies AC voltage at two adjacent frequencies, exploiting the different frequency dependencies of ferroelectric switching current (frequency-independent), dielectric displacement current (linearly frequency-dependent), and ohmic leakage current (frequency-independent but voltage-dependent)[99]. This technique enables the extraction of genuine hysteresis loops without requiring high DC field stress that could damage samples. The PUND technique, adapted for nanoscale measurements using AFM tips as electrodes, provides another powerful approach for separating displacement currents from leakage[100]. This method applies sequences of voltage pulses and monitors current responses during and between successive pulses, enabling the identification of polarization switching-related currents versus background leakage. For elementary ferroelectrics measured via PFM, frequency-dependent amplitude sweeps can discriminate genuine piezoelectric/ferroelectric responses from artifacts like Vegard strain or electrochemical effects[101]. True ferroelectric responses show minimal frequency dependence within the measurement bandwidth, while artifacts often exhibit strong frequency dependence.

The piezoelectric effect, where mechanical stress generates electric polarization, is closely linked to ferroelectricity. Elemental ferroelectrics exhibit unique piezoelectric properties, often differing from conventional compound ferroelectrics. An important discovery is the giant negative longitudinal piezoelectric response in 2D Group-Va monolayers[89]. This counterintuitive effect, where polarization decreases under tensile strain, contrasts with the positive response in conventional materials. This anomalous response is attributed to the buckling-driven polarization mechanism. First-principles calculations reveal a pronounced nonanalytic behavior in the piezoelectric response, with different coefficients under tensile and compressive strains, highlighting the complex electromechanical coupling. In Bi monolayers, the internal strain contribution to the piezoelectric response is exceptionally strong, dominating the clamped-ion response and highlighting the role of structural distortions, as shown in Figure 3. This dominance of internal strain is a key factor in the negative piezoelectric response. These unique properties open possibilities for novel electromechanical devices with functionalities not achievable with conventional materials, enabling new sensor technologies and actuator designs.

The discovery of intrinsic ferroelectricity in single-element materials has generated significant interest due to their potential applications. The advantages of ultrathin ferroelectric devices, leveraging the 2D nature of many elemental ferroelectrics, are compelling for next-generation nanoelectronics[21,15]. Te nanowires, with room-temperature ferroelectricity, high carrier mobility, and resistive switching, are promising for high-density data storage, potentially exceeding terabytes per square centimeter[23]. Their resistive switching also enables self-gated ferroelectric field-effect transistors (SF-FETs) with nonvolatile memory, potentially advancing low-power electronic circuits. Bi monolayers, with charged domain walls and topologically protected interfacial states, hold promise for nanoelectronics and spintronics, where domain manipulation and topological edge states could lead to novel functionalities[91,97]. The control of domain walls and the electronic properties of interfacial states open possibilities for nanoscale devices like domain-wall-based memory and spin-based transistors. Te multilayers, with tunable polarization and spin textures from spin-orbit coupling, offer a platform for spintronic devices, enabling spin-based memory and logic with enhanced efficiency[21]. The discovery of sliding ferroelectricity in multilayer graphene further expands applications, presenting opportunities for multi-state memory and memristive switching. However, realizing these applications requires overcoming challenges related to material stability, scalability, and integration with existing semiconductor manufacturing.

In conclusion, intrinsic ferroelectricity in single-element materials represents an important shift in our understanding of ferroic phenomena. The mechanisms governing polarization and switching, linked to structural distortions, offer possibilities for novel electronic and spintronic devices. Continued research, focusing on material discovery, device optimization, and technological translation, will be essential for realizing the important potential of elementary ferroelectrics.

## 2.3. Stacking ferroelectricity

Stacking ferroelectricity represents a novel approach to realize ferroelectricity in 2D van der Waals (vdW) materials, as shown in Figure 4, where ferroelectric properties arise not from the intrinsic properties of individual monolayers, but from the way these layers are stacked together, first introduced by Li and Wu[102]. This concept is particularly significant because it allows for the creation of ferroelectric behavior even from non-ferroelectric monolayers by breaking the inversion symmetry through specific stacking configurations. Different from conventional ferroelectrics, where polarization originates from ionic displacement within a unit cell, stacking ferroelectricity is rooted in the interlayer charge compensation and the resultant dipole formation between adjacent layers. This mechanism expands the family of 2D ferroelectrics and offers new possibilities for designing and controlling ferroelectric properties at the nanoscale.

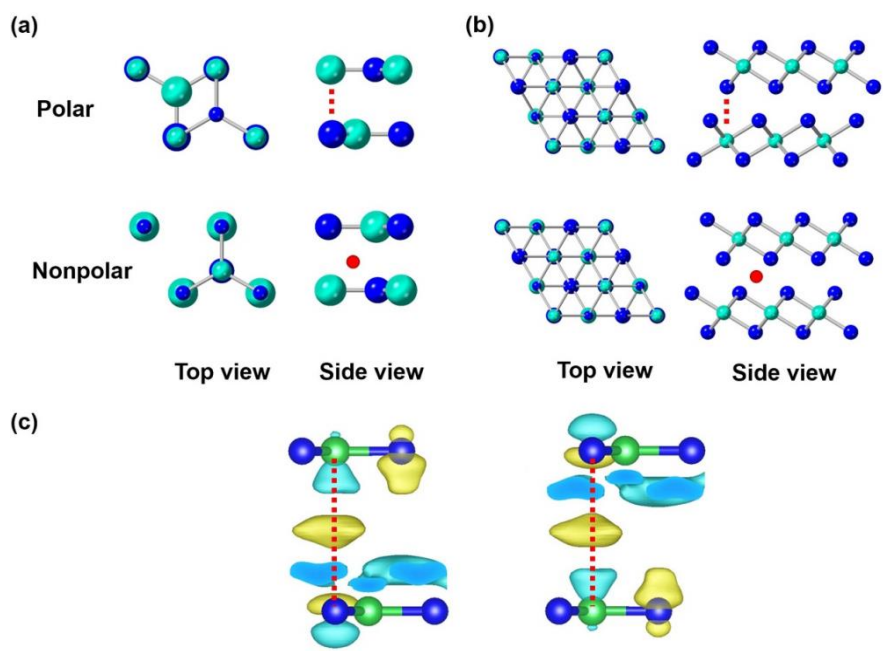

**Figure 4.** Polar stacking versus nonpolar stacking. (a) The structure of the bilayers same as BN. (b) The structure of the bilayers same as T-WS2. Different colors are used to represent different atom types. The sizes of the atoms of the upper layers are reduced in the top view. Red dashed lines denote the unequal Coulomb interaction due to the stacking. The red solid circle represents the spatial inversion symmetry. (c) Differential charge density for polar stacking bilayer BN with downward (left) and upward (right) OP polarization. Green (blue) spheres represent B (N) atoms. The yellow (cyan) color indicates regions of electron accumulation (depletion). Reproduced with permission.[27] Copyright 2025, IOP Publishing.

The theoretical framework for understanding stacking ferroelectricity involves group theory analysis, which is crucial for predicting and explaining the emergence of polarization in stacked vdW materials[103]. Group theory helps to identify the symmetry operations that are broken or preserved upon stacking, thereby determining whether a bilayer or multilayer structure can exhibit ferroelectricity. A key aspect of this theory is the classification of symmetry operations into those that can and cannot exchange the positions of stacked layers, denoted as $R_-$ and $R_+$, respectively. The presence of an $R_-$ symmetry, such as inversion symmetry, in the bilayer leads to the absence of out-of-plane polarization. The stacking operation, which describes how one monolayer is positioned relative to another, is crucial in determining the symmetry of the resulting bilayer. It is represented as $\tau_z O$, where $\tau_z$ is a trivial out-of-plane translation and $O$ is an in-plane transformation operator that includes rotational and translational components. By analyzing the layer group of a monolayer and how it transforms under different stacking operations, one can predict the polar properties of the bilayer. This approach not only explains previously observed sliding ferroelectricity but also provides a generalized method for designing new ferroelectric bilayers.

The origin of polarization in stacking ferroelectrics can be understood from both classical and quantum perspectives. Classically, the polarization arises from the distortion of electron clouds and the formation of dipoles due to uncompensated Coulomb interactions between layers. Quantum mechanically, the Berry phase theory provides a framework to calculate the electric polarization by considering the momentum space and symmetry of the crystal structure when dealing with 2D out-of-plane polarization[29]. In essence, the breaking of inversion symmetry upon stacking, combined with interlayer charge transfer, is the fundamental mechanism driving stacking ferroelectricity.

Experimental verification of stacking ferroelectricity has been achieved in various vdW material systems, highlighting the versatility of this approach[25,26,50,104–111]. Bilayer hexagonal boron nitride (h-BN) was one of the first systems where sliding ferroelectricity was experimentally confirmed[26]. In bilayer h-BN, the polar stacking configuration, such as AB stacking, breaks the inversion symmetry present in the individual monolayers, leading to out-of-plane ferroelectricity. The polarization in bilayer h-BN is switchable by interlayer sliding, which can be induced by applying an electric field or mechanical force[25].

Beyond h-BN, stacking ferroelectricity has been demonstrated in transition metal dichalcogenides (TMDs) and other layered materials. In bilayer $WTe_2$, sliding ferroelectricity was observed through transport measurements, indicating a metal-ferroelectric behavior[3]. Recent studies have also explored stacking ferroelectricity in heterostructures and multilayers.

In twisted bilayer graphene, ferroelectric domains and domain walls have been observed, indicating the emergence of ferroelectric order from stacking-induced symmetry breaking[108]. In multilayer systems like 3R-MoS$_2$, multiple polarization states have been identified, attributed to the stacking-dependent interlayer interactions[50]. These experimental realizations across diverse material systems underscore the broad applicability of stacking ferroelectricity and its potential for creating novel electronic properties and functionalities.

In conclusion, stacking ferroelectricity represents a significant advancement in the field of 2D materials, offering a versatile platform for creating novel electronic properties and device functionalities. The ongoing research efforts and future directions hold great promise for realizing the full potential of stacking ferroelectrics in next-generation nanoelectronics and beyond.

**2.4. Polar metallicity**

The coexistence of ferroelectricity and metallicity poses a fundamental challenge in condensed matter physics. The conventional understanding dictates that itinerant electrons in metals should effectively screen the long-range Coulomb interactions that are considered indispensable for stabilizing spontaneous polarization. This apparent contradiction was first theoretically addressed by Anderson and Blount in their seminal work, where they predicted the theoretical possibility of symmetry-breaking structural transitions within metallic systems[112]. However, experimental confirmation remained elusive for nearly five decades. This changed with the discovery of LiOsO$_3$ by Shi *et al.* in 2013[30], which provided the first experimental realization of a continuous centrosymmetric-to-polar structural transition at 140 K in a material that maintained metallic conductivity. This discovery established a new class of materials termed ferroelectric-like metals. The seminal work on LiOsO$_3$ revealed a displacive transition mechanism, remarkably similar to the ferroelectric transition observed in archetypal ferroelectric insulators such as LiNbO$_3$[113]. This transition involves a subtle shift of lithium ions along the $c$-axis by approximately 0.5 Å, occurring without disrupting the charge delocalization inherent to the metallic Os-O network[30]. It undergoes a second-order structural phase transition from a centrosymmetric to a non-centrosymmetric phase, exhibiting a polar distortion while retaining metallic conductivity. However, due to the presence of itinerant electrons that screen internal electric fields, the polarization in bulk LiOsO$_3$ is not switchable, and therefore it does not fulfill the conventional criteria for ferroelectricity. Theoretical investigations demonstrate that ferroelectricity can be achieved in ultrathin LiOsO$_3$ films, where two-unit-cell-thick films exhibit switchable polarization through asymmetric hysteresis loops, as surface Li ions experience incomplete screening compared to fully screened internal ions[114]. The 2-UC LiOsO$_3$ film remains metallic but possesses a partially switchable electric dipole moment under an external electric field. This finding directly challenged the conventional notion that metallic screening would inevitably suppress such ionic displacements, thereby reigniting significant interest and activity in the field of polar metallicity. Subsequent theoretical investigations by Benedek and Birol[115] further elucidated the conditions conducive to polar metallicity, identifying two crucial requirements: first, a weak coupling between conduction electrons and the polar phonons associated with the ferroelectric instability, and second, primary structural instabilities driven by geometric factors rather than purely electronic effects[115].

While the coexistence of metallicity and polar distortions has been demonstrated in various systems, a critical distinction must be made between true "metallic ferroelectrics" and "polar metals". The term "ferroelectric" inherently requires switchable polarization, yet in most metallic systems, itinerant electrons effectively screen internal electric fields, preventing polarization reversal. Consequently, only a remarkably limited number of systems can genuinely be classified as metallic ferroelectrics. Experimentally, the landmark achievement belongs to 2D WTe$_2$, where Fei et al. demonstrated bistable ferroelectric switching in few-layer films through gate-controlled measurements, with switching capability persisting up to 350 K[3]. This represents the first unambiguous experimental verification of switchable polarization in a metallic system. Theoretically, several predictions have emerged: Filippetti et al. predicted native ferroelectricity in layered Bi$_5$Ti$_5$O$_{17}$, showing that despite metallicity, this material can sustain switchable polarization of ~35 μC/cm² due to its unique self-screening mechanism[116]. Lu et al. demonstrated that ultrathin LiOsO$_3$ films (2-unit cells) exhibit asymmetric ferroelectric hysteresis, where surface Li ions remain switchable while internal atoms are screened[114]. Additionally, Luo et al. proposed 2D CrN and CrB$_2$ as hyperferroelectric metals, where out-of-plane polarization can be switched despite in-plane metallicity[117]. In contrast, bulk systems like LiOsO$_3$, Cd$_2$Re$_2$O$_7$, and doped perovskites represent polar metals—materials with broken inversion symmetry but lacking switchable polarization due to complete electronic screening. This fundamental distinction highlights the rarity of true metallic ferroelectricity and underscores the exceptional nature of the few confirmed cases.

Established methods for creating conducting polar materials have traditionally relied on doping strategies[118]. In ferroelectric oxides like BaTiO$_3$, introducing mobile electrons via mechanisms such as oxygen vacancy formation has been a common approach[119]. At low concentrations, these dopants provide charge carriers without quenching the material's ferroelectric properties.

A more recent and widely studied approach involves engineering polar conducting interfaces, most notably the LaAlO$_3$/SrTiO$_3$ (LAO/STO) heterostructure[120–123]. This system hosts a high-mobility two-dimensional electron gas (2DEG) at the interface between the two insulating oxides[121,124]. The formation of this 2DEG is driven by an electronic reconstruction known as the "polar catastrophe"[121,125]. When the polar LAO layer is grown on the non-polar STO substrate, an electrostatic potential builds up. Above a critical thickness of LAO, this potential is averted by the transfer of electrons to the interface, creating the conductive 2DEG. The LAO/STO interface is a rich platform for emergent phenomena, where gate-tunable superconductivity and magnetism can coexist with the polarity of the LAO layer. The conductivity of this interface can be controlled by external stimuli like electric fields[125,126]. The success of this system has spurred research into other polar/non-polar interfaces, demonstrating that the polar catastrophe mechanism is a general principle for generating interface conductivity[124,127]. The properties of the resulting 2DEG can also be tuned by changing the substrate orientation[128].

These classical doping and interface engineering approaches are crucial for understanding the more exotic intrinsic polar metals. They demonstrate that electronic screening and polar ordering are not mutually exclusive. The LAO/STO system, in particular, bridges the gap between conventional doped ferroelectrics and intrinsic polar metals, showcasing how confinement and interfacial effects can stabilize conducting polar states.

Recent advancements have significantly broadened the landscape of polar metals, revealing that the coexistence of polarity and metallicity can arise through a variety of mechanisms. These mechanisms can be conceptually understood by considering the total polarization ($P$) as a sum of contributions from different sources[129]:

$$P = P_{\text{ionic}} + P_{\text{electronic}} + P_{\text{geometric}} \qquad (1)$$

In metallic systems, the geometric polarization ($P_{\text{geometric}}$) becomes increasingly crucial[129], arising from mechanisms such as:

1. *Hybrid Improper Ferroelectricity*: Demonstrated in Ruddlesden-Popper phases like Ca$_3$Ru$_3$O$_7$[130], where trilinear coupling between non-polar octahedral rotation modes ($a^0a^0c^-$ tilt) and in-phase tilts ($a^0a^0c^+$) generates net polarization.
2. *Geometric Frustration*: Observed in pyrochlore Cd$_2$Re$_2$O$_7$[131], where competing domain configurations produce polarization.

3. *Van der Waals Layering*: Realized in 2D WTe$_2$[3], where interlayer sliding creates switchable polarization without breaking in-plane metallicity[51]

First-principles computational methods, particularly DFT, have played an indispensable role in elucidating the underlying physics of polar metallicity and guiding the search for new materials. Zhao *et al.*[132] demonstrated through DFT calculations that certain crystal symmetries can permit partial retention of polarization even in metallic systems by considering the contributions to the inverse high-frequency dielectric constant ($\epsilon_\infty^{-1}$) [132]

$$\epsilon_\infty^{-1} = \epsilon_{total}^{-1} - \epsilon_{interatomic}^{-1} \qquad (2)$$

This equation highlights that materials exhibiting strong interatomic polarizabilities ($\epsilon_{interatomic}$) can sustain a finite polarization even when the total dielectric constant ($\epsilon_{total}$) diverges in the metallic state due to electronic screening.

Despite significant progress, achieving fully switchable polarization in bulk polar metals remains a considerable challenge. The discovery of reversible polarization in 2D WTe$_2$ marked a breakthrough, demonstrating that quantum confinement in van der Waals materials can indeed enable electric-field switching in a metallic system[3]. This concept was later extended to twisted bilayer graphene systems, where moiré potentials can create polarization-mutable quantum dots, offering another pathway to control polarization in metallic environments[133].

Also, the balance between remanent polarization and metallic conductivity remains a central challenge. In conventional ferroelectrics, a large remanent polarization arises from robust off-center ionic displacements that are stabilized by long-range Coulomb interactions. However, in metallic systems, free carriers tend to screen internal electric fields, thereby weakening the driving force for such displacements. As a result, increasing the carrier concentration to enhance conductivity often comes at the expense of the structural distortions necessary for a sizable polarization. Conversely, strengthening polar displacements to maximize remanent polarization typically reduces the bandwidth or carrier mobility, suppressing metallicity. This competition imposes a fundamental constraint on simultaneously optimizing both properties. Therefore, attaining equilibrium between substantial remanent polarization and robust metallic conductivity continues to be a pivotal hurdle in the advancement of functional polar metals.

The intricate interplay between superconductivity and polar metallicity has also emerged as a particularly intriguing area of research. Rischau *et al.* observed enhanced superconductivity in the vicinity of a ferroelectric quantum critical point in SrTiO$_3$-based systems, suggesting that soft polar modes might play a role in mediating unconventional Cooper pairing[4]. Theoretical work by Volkov et al.[134] proposed a microscopic mechanism for this enhancement, highlighting the cooperative effects of transverse optical phonon frequency ($\omega_{TO}$), electron-phonon coupling ($\lambda_{ep}$), and electron-electron coupling ($\lambda_{ee}$) in boosting the superconducting transition temperature ($T_c$):

$$T_c \propto \omega_{TO} \exp\left(-\frac{1}{\lambda_{ep} + \lambda_{ee}}\right) \quad (3)$$

Materials design strategies for polar metals continue to evolve, guided by computational approaches and a deeper understanding of the underlying mechanisms. Hickox-Young et al. proposed a polar metals taxonomy classifying known compounds into distinct types based on the origin of their polar metallic state[135]:

- Type I: Proper ferroelectrics with weak screening (e.g., doped $BaTiO_3$)
- Type II: Improper geometric ferroelectrics (e.g., $LiOsO_3$)
- Type III: Hybrid systems with coupled order parameters (e.g., $Ca_3Ru_2O_7$)

Recent experimental breakthroughs further highlight the expanding possibilities of this field, including the discovery of room-temperature polar metals in strained La-doped $BaTiO_3$ films[136] and the observation of anisotropic polarization-dependent conductivity in $KNbO_3/BaTiO_3$ superlattices[137]. Meanwhile, fundamental questions remain regarding the ultimate limits of polarization magnitude in metals and the possibility of multiferroic metallic systems. The prediction of altermagnetic-polar metal coupling in h-$YMnO_3$[138] suggests new avenues for spintronic applications. Concurrently, advances in oxide molecular beam epitaxy have enabled atomic-scale engineering of polar metallic interfaces[139], opening up new possibilities for creating artificial polar metallic structures with tailored properties.

The field of polar metallicity is now poised at the threshold of technological applications. Proposed devices range from non-volatile superconducting memory elements based on $NbSe_2/In_2Se_3$ heterostructures[140] to topological quantum bits utilizing polar metal Josephson junctions[141] and ultra-low power electronics leveraging gate-tunable polarization in 2D metals[51]. Future research directions must address critical challenges in materials synthesis, particularly in stabilizing switchable polarization in three-dimensional metals and achieving a comprehensive understanding of the dynamics of polar domains within conducting matrices.

### 2.5. Fractional quantum ferroelectricity

Recently, Ji et al.[17] introduced the concept of fractional quantum ferroelectricity (FQFE), in which atomic displacements correspond to a fractional multiple of the lattice constant. Before delving into the details of FQFE, it is instructive to briefly review the characteristics of conventional ferroelectrics and quantum ferroelectricity (QFE)[17,142] as a basis for comparison. In conventional ferroelectrics (e.g., $BaTiO_3$ and $PbTiO_3$), the spontaneous polarization arises from small atomic displacements, resulting in the well-known form of ferroelectricity illustrated in Figure 5a. In such cases, the polarization direction is consistent with the symmetry of the ferroelectric phase, in accordance with Neumann's principle[143], which states that the symmetry elements of any physical property of a crystal must include all the symmetry elements of the crystal's point group. In QFE (Figure 5b), the anions or cations are displaced by an integer multiple of the lattice vector, producing an integer-quantized polarization[142]. For FQFE, the ion displaces a fractional lattice vector (e.g., -0.5 a + 0.5 b in the case of square lattice, see Figure 5c).

To clarify more clearly the essential features of FQFE, we consider a hexagonal system as a representative example. Here, the ferroelectric behavior manifests in two distinct low-symmetry phases, denoted L1 and L2 (Figures 5d and 5f), both adopting a $MoS_2$-type structure. These phases belong to space group P-6m2 with point group $D_{3h}$, which is a non-polar point group. Each structure comprises both M and F ions (Figure 5d), with the M ions being mobile and responsible for polarization switching, while the F ions remain fixed. For conceptual clarity, an intermediate high-symmetry paraelectric phase, H (Figure 5e), is introduced. This phase has the same chemical composition as L1 and L2. The L1 and L2 configurations can be generated from H by shifting the M atom along the blue and red arrows, respectively (Figure 5e), leading to a net dipole moment difference between the two phases (green double-headed arrow). If the energy barrier between L1 and L2 is sufficiently small, an external electric field can switch between them, thereby realizing ferroelectricity. Within the framework of the modern theory of polarization (MTP), both the absolute polarization and the polarization difference between L1 and L2 are multivalued quantities. A schematic lattice representation of these dipole moments is shown in Figure 5g, where the blue and red spheres denote the dipole moments of L1 and L2, respectively, and the green double-headed arrows represent their difference. Notably, this polarization difference retains a finite in-plane component that is not invariant under certain symmetry operations, such as the threefold rotation about the z-axis ($C_{3z}$), that characterize the low-symmetry phases L1 and L2. Once the transformation path between L1 and L2 is determined, the polarization difference between them is also determined.

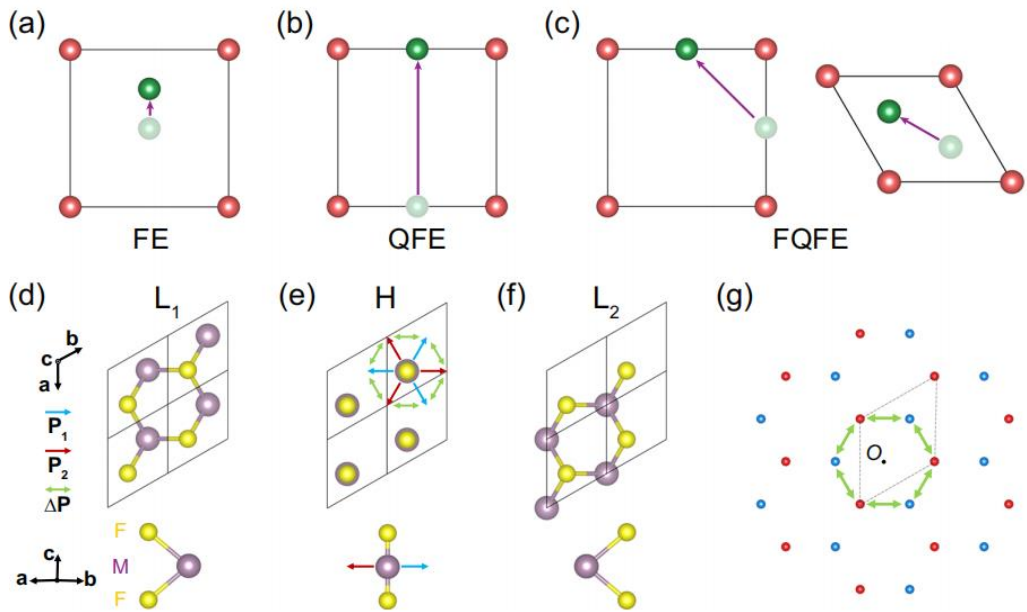

**Figure 5.** Concept of FQFE. Schematics of (a) FE, (b) QFE, and (c) FQFE, supposing the ion with ±1 charges. The green and red balls represent movable ions and ligand ions, respectively. Top and side views of the FQFE example: (d) low-symmetry phase $L_1$, (e) high-symmetry phase H, (f) low-symmetry phase $L_2$. The black border indicates the unit cell. F includes two fixed atoms (layers) F, and M is a movable atom (layer). The blue and red arrows depict the atomic displacements of M from H to $L_1$ and $L_2$, respectively. The green arrows show ΔP, the atomic displacements of M between $L_1$ and $L_2$. Since only M moves, the blue, red, and green arrows can also represent P1 (polarization of L1), P2 (polarization of $L_2$), and ΔP (polarization difference between low symmetry phases), respectively. ΔP cannot be invariant under a point symmetry operation ($C_{3z}$) of the low-symmetry phase, which leads to the FQFE. (g) The latticed form of $P_1$, $P_2$, and ΔP. The black dashed parallelogram depicts the "lattice" of polarization. The blue and red points represent $P_1$ and $P_2$, respectively. ΔP can be any vector between points with different colors. Therefore, ΔP is non-zero and fractionally quantized, i.e., 1/3 Q along the [120] direction. Reproduced with permission.[27] Copyright 2023, IOP Springer Nature Limited.

A defining feature of FQFE is its violation of Neumann's principle. The direction of polarization in FQFE is not constrained by the symmetry of the polar phase, challenging the conventional symmetry-based understanding of ferroelectricity. Group theory analysis plays a crucial role in identifying materials exhibiting FQFE. As demonstrated by Yu *et al.* [31], a symmetry strategy can be employed to efficiently screen for FQFE candidates. This strategy involves analyzing the space groups of crystal structures and identifying symmetry operations that can lead to symmetrically equivalent phases with fractional atomic displacements. Through high-throughput screening combined with first-principles calculations, a large number of potential FQFE materials have been identified, significantly expanding the known pool of ferroelectrics.

Monolayer $\alpha$-In$_2$Se$_3$ serves as a prominent example of FQFE, exhibiting an unexpected in-plane polarization despite its $P3m1$ symmetry, which should only permit out-of-plane polarization according to Neumann's principle[144–146]. Ji et al.[31] have shown that the in-plane polarization in $\alpha$-In$_2$Se$_3$ arises from FQFE, with Se atoms in the middle layer undergoing fractional displacements, leading to a polarization difference that is a fractional multiple of the polarization quantum. First-principles calculations and symmetry analysis have successfully explained this phenomenon, demonstrating the power of the FQFE theory in understanding unconventional ferroelectric behaviors discovered experimentally [147,148]. The switching barrier in monolayer $\alpha$-In$_2$Se$_3$ is calculated to be around 40 meV/f.u.[149], confirming that the FQFE can indeed be realized experimentally.

Beyond 2D materials, FQFE is also predicted to exist in bulk materials. Unlike conventional ferroelectrics, FQFE systems can exhibit polarization values that are fractional or integer multiples of this quantum, leading to type-I and type-II FQFE[31]. Bulk AgBr, which crystallizes in the non-polar zinc blende structure with $F\bar{4}3m$ space group, is theoretically shown to exhibit type-I FQFE[31]. Although AgBr's ground state is a rocksalt structure with $Fm\bar{3}m$ symmetry, the $F\bar{4}3m$ phase is dynamically stable and energetically close, allowing for FQFE to emerge via fractional displacement of Br atoms. The calculated switching barrier for AgBr is remarkably low, around 22 meV/f.u. for in-plane switching and 155 meV/f.u. for out-of-plane switching, indicating its potential for low-power ferroelectric devices. Furthermore, type-II FQFE with an integer polarization quantum but fractional lattice vector displacement has been demonstrated in monolayer HgI$_2$, which possesses a non-polar $P\bar{4}m2$ space group. Despite the non-polar point group, HgI$_2$ exhibits a large spontaneous polarization of 42 $\mu$C/cm$^2$ due to fractional displacement of Hg atoms, with a polarization difference equal to the polarization quantum. The switching barrier in HgI$_2$ is calculated to be 68 meV/f.u., further highlighting the possibility of realizing low-barrier switching in FQFE materials.

We compare the FQFE to electrets, which have permanent polarization due to oriented dipoles. Electrets are dielectric materials that exhibit a quasi-permanent electric charge or dipole polarization, often due to trapped real charges, aligned molecular dipoles, or frozen-in ionic displacements, but this polarization is not necessarily coupled to a crystallographic phase transition or switchable between crystallographically equivalent states in the same way as in ferroelectrics. While both FQFE materials and electrets can exhibit macroscopic polarization, the underlying physical origins and switching mechanisms are fundamentally different. The ion displacement in FQFE is constrained by symmetry (i.e., from one high symmetry Wyckoff position to another high symmetry Wyckoff position), unlike the statistical ion migration in ion conductors.

The existence of FQFE expands the scope of ferroelectricity beyond the traditional constraints of polar point groups and small atomic displacements. It challenges the conventional understanding that polarization direction must conform to the symmetry of the polar phase (Neumann's principle), as FQFE systems can exhibit polarization forbidden by its point group. It opens new avenues for exploring novel ferroelectric materials with unique properties, including potentially high polarizations and low switching barriers. The theoretical framework of FQFE, combined with high-throughput computational screening, provides a powerful approach for discovering and designing next-generation ferroelectric materials for diverse applications in nanoelectronics and spintronics. The interplay between FQFE and other ferroic orders, as well as the switchability and broader range of FQFE candidates, remains an exciting direction for future research. Future research in FQFE will likely focus on several key areas. Experimental verification of FQFE in a wider range of predicted candidate materials, beyond α-$In_2Se_3$, is crucial. Understanding the detailed switching dynamics, including the role of kinetic barriers, will be important for device applications.

## 2.6. Wurtzite-Based Ferroelectrics

Wurtzite-based ferroelectrics represent an important advancement in the field of ferroelectric materials, offering opportunities for seamless integration with conventional semiconductor technologies[32–34]. Unlike traditional ferroelectric oxides such as perovskites, wurtzite ferroelectrics are characterized by their tetrahedral coordination and direct compatibility with silicon-based CMOS processing[60,150]. The discovery of ferroelectricity in doped aluminum nitride (AlN) by Fichtner et al. in 2019 marked a watershed moment, demonstrating robust ferroelectric switching in scandium-doped aluminum nitride ($Al_{1-x}Sc_xN$) at room temperature[32]. This breakthrough has catalyzed intensive research into wurtzite-structured III-nitride compounds, positioning them as promising candidates for next-generation memory devices and neuromorphic computing applications[151,152,33,153,154].

The fundamental crystal structure of wurtzite ferroelectrics consists of tetrahedrally coordinated cations surrounded by four anions, shown in Figure 6, arranged in a hexagonal close-packed lattice with space group *$P6_3mc$*. This structural framework inherently possesses a polar axis along the c-direction, creating spontaneous polarization even in binary compounds like aluminum nitride[155]. However, pristine wurtzite materials typically exhibit prohibitively high coercive fields that preclude practical ferroelectric switching[156]. The key to realizing switchable ferroelectricity lies in strategic chemical doping, which modifies the electronic structure and reduces the energy barriers for polarization reversal[155,157]. In $Al_{1-x}Sc_xN$ alloys, scandium substitution on aluminum sites introduces larger ionic radii and modified bonding characteristics that stabilize intermediate polar configurations during switching, effectively lowering the coercive field from unattainable values in pure AlN to practically manageable ranges of 1.5-6.5 MV/cm[158].

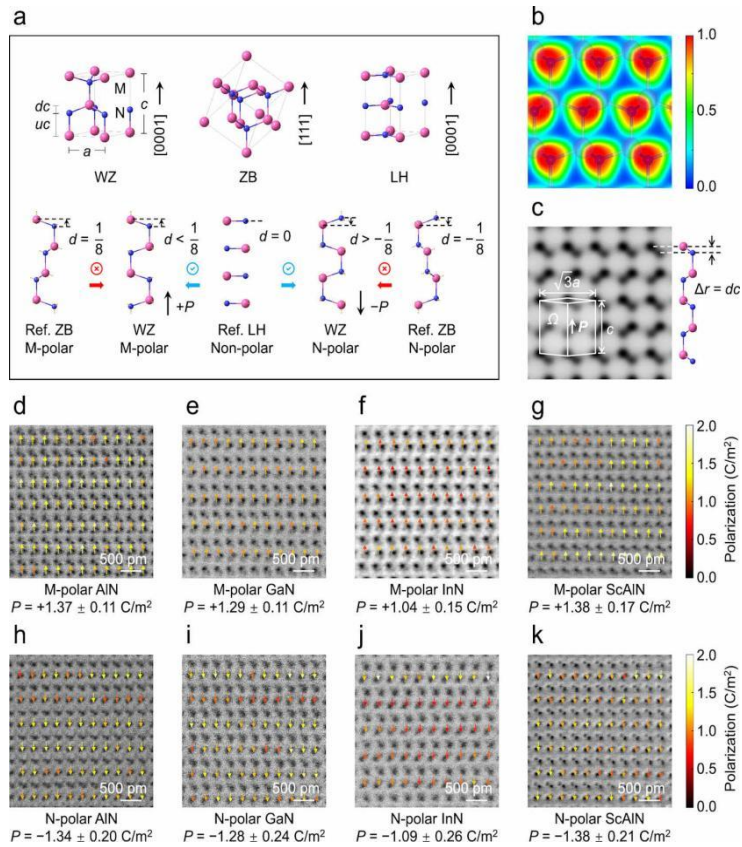

**Figure 6.** (a) Crystal structures of III-nitrides, including wurtzite (WZ), zinc-blende (ZB), and a h-BN-like layered-hexagonal (LH) configurations (top), and atomic schematics illustrating the adiabatic and gap-preserving displacement between WZ and its ZB and LH reference structures along the c-axis (bottom). (b) Electron localization function (ELF) of WZ AlN projected along the $[11\bar{2}0]$ zone-axis, showing both the ionic ($P_{ion}$) and electronic ($P_{el}$) contributions to polarization. (c) Diagram of the local polarization measurement method based on a simulated ABF-STEM image of WZ AlN. Atomic displacement ($\Delta r$) and lattice constants (a and c) are measurable. The inset shows a quadrilateral prism representing the unit cell with a volume of $\Omega$. (d)–(k) ABF-STEM images of WZ M-polar (d) AlN, (e) GaN, (f) InN, and (g)

ScAlN, and N-polar (h) AlN, (i) GaN, (j) InN, and (k) ScAlN, each overlaid with vector map of the measured local polarization. Reproduced with permission from [159] Copyright 2025, Springer Nature Limited.

The ferroelectric switching mechanisms in wurtzite materials exhibit fascinating complexity that deviates significantly from conventional ferroelectric behavior[160]. Research has revealed two distinct switching pathways: collective and individual mechanisms[160]. The collective mechanism, observed in binary wurtzite compounds and low-doping-concentration alloys, involves the simultaneous displacement of entire cation and anion sublattices through a nonpolar hexagonal intermediate structure (wz$^+$ → h$^0$ → wz$^-$). In contrast, the individual switching mechanism emerges in heavily doped alloys and multinary wurtzite compounds, where ferroelectric reversal proceeds through sequential inversion of individual tetrahedral units. This individual pathway is characterized by multiple energy minima along the switching trajectory, creating a fractal-like domain wall structure that enables faster switching kinetics. Remarkably, the transition from collective to individual switching occurs at a critical scandium concentration of approximately x = 0.22-0.28 in $Al_{1-x}Sc_xN$, coinciding with a dramatic reduction in switching barriers. The individual mechanism involves the breaking and reformation of cation-anion bonds in a cascade-like manner, where a single aluminum ion displacement triggers a columnar propagation of switched tetrahedra, fundamentally altering our understanding of domain wall dynamics in ferroelectrics.

Material diversity within the wurtzite ferroelectric family continues to expand rapidly through systematic compositional engineering[155,157]. Beyond the pioneering $Al_{1-x}Sc_xN$ system, researchers have successfully demonstrated ferroelectricity in aluminum boron nitride ($Al_{1-x}B_xN$), aluminum yttrium nitride ($Al_{1-x}Y_xN$), and aluminum gadolinium nitride ($Al_{1-x}Gd_xN$) alloys[155,154,157]. Each composition offers unique advantages and trade-offs in terms of polarization magnitude, coercive field, and thermal stability[150]. $Al_{1-x}Sc_xN$ exhibits the largest remanent polarization values, reaching up to 165 $\mu C/cm^2$, while $Al_{1-x}B_xN$ demonstrates superior thermal stability at elevated temperatures[150,161]. The recent computational prediction and experimental verification of $Al_{1-x}Gd_xN$ ferroelectricity represents a significant milestone, as it introduces magnetic rare-earth elements into wurtzite ferroelectrics, potentially enabling multiferroic functionality[157]. First-principles calculations indicate that the minimum energy pathway for switching in these rare-earth-doped systems changes from collective to individual processes at gadolinium fractions exceeding 10-15%, consistent with the switching mechanisms observed in other doped wurtzite alloys. Jiang et al. demonstrated that epitaxial AlN/ScN superlattices exhibit strain-induced multifunctionality, with four distinct strain regions enabling polarization control from zero to 1.07 C/m² in nominally paraelectric materials. This approach showcases giant electro-optic coefficients and tunable band gaps spanning blue to ultraviolet wavelengths, highlighting superlattice engineering's potential for wurtzite ferroelectrics.

Wurtzite ferroelectrics possess several distinctive properties that distinguish them from conventional ferroelectric materials[158]. Their wide bandgaps (4.9-5.6 eV) ensure low leakage currents and high breakdown fields, crucial for reliable device operation. The exceptional thermal stability, with some compositions maintaining ferroelectric properties up to 1100°C, enables operation in harsh environments where traditional ferroelectrics would fail. Perhaps most importantly, their compatibility with standard semiconductor processing techniques, including low-temperature deposition on flexible substrates, opens unprecedented possibilities for integration into existing CMOS fabrication lines[60]. The ability to achieve substantial polarization values while maintaining CMOS compatibility represents a paradigm shift in ferroelectric device technology.

Despite their promising attributes, wurtzite ferroelectrics face several technical challenges that require continued research attention[152,162]. The high coercive fields, while reduced compared to pure materials, still approach breakdown limits and necessitate careful engineering of switching pulse protocols. Cycling endurance remains a concern, with fatigue mechanisms linked to defect generation during repeated polarization switching[152,163]. Compositional segregation in alloy systems can lead to non-uniform properties and switching behavior, requiring precise control over film deposition and annealing conditions. Additionally, the complex relationship between local chemical environment and switching barriers demands sophisticated theoretical modeling to guide materials optimization[155,164].

Future research directions for wurtzite ferroelectrics encompass both fundamental understanding and technological advancement. Theoretical efforts focus on developing comprehensive models of individual switching mechanisms and their relationship to local chemical environments, potentially guided by machine learning (ML) approaches for materials optimization. Experimental priorities include extending compositional space through exploration of quaternary and higher-order alloys, investigating novel deposition techniques for improved film quality, and developing interface engineering strategies to enhance device performance. The integration of wurtzite ferroelectrics with emerging technologies such as flexible electronics, biocompatible implants, and quantum devices presents exciting opportunities for next-generation applications. As our understanding of switching mechanisms deepens and processing techniques mature, wurtzite-based ferroelectrics are poised to play an increasingly important role in the evolution of electronic devices, offering a pathway to bridge the gap between traditional ferroelectric functionality and modern semiconductor technology requirements.

### 2.7. Freestanding Ferroelectric Membranes

Freestanding ferroelectric membranes are another advancement in ferroelectric materials research, offering unprecedented opportunities to explore and manipulate ferroelectric properties by liberating these materials from the constraints imposed by rigid substrates[35–38]. Unlike conventional epitaxial thin films that are permanently anchored to their growth substrates, freestanding membranes can be mechanically transferred to arbitrary platforms, enabling the exploration of strain-tunable properties and novel functionalities that are inaccessible in substrate-constrained systems[165,166]. As shown in Figure 7, this freedom from epitaxial constraints has opened new avenues for investigating the fundamental physics of ferroelectricity while simultaneously providing a versatile platform for next-generation flexible electronic devices.

The concept of freestanding ferroelectric membranes emerged from the recognition that substrate clamping significantly restricts the electromechanical response of ferroelectric materials. In epitaxial thin films, the rigid substrate prevents the free expansion and contraction of the ferroelectric layer, thereby suppressing piezoelectric coefficients and limiting domain dynamics. By releasing these films from their substrates, researchers have discovered that ferroelectric membranes exhibit dramatically enhanced piezoelectric responses, with coefficients that can exceed those of bulk materials by factors of 2-3[37]. This enhancement arises from the elimination of substrate clamping effects and the introduction of new mechanical degrees of freedom that allow for more complex domain configurations and polarization switching pathways[167]. Theoretical studies by Lu et al. revealed novel ferroelectric mechanisms in ultrathin perovskite oxide films, including surface-effect induced ferroelectricity that strengthens with decreasing thickness[168]. These findings provide crucial insights into the fundamental physics governing freestanding membranes, particularly regarding surface-dominated effects and thickness-dependent ferroelectric enhancement in substrate-free configurations.

The fabrication of freestanding ferroelectric membranes primarily relies on water-soluble sacrificial layer techniques, which have been refined and optimized over the past decade[35,166]. The most commonly employed approach utilizes $Sr_3Al_2O_6$ (SAO) as a sacrificial buffer layer, which is epitaxially grown on perovskite substrates such as $SrTiO_3$ before the deposition of the ferroelectric film[35,169]. The high water solubility of SAO enables selective etching in aqueous solutions, allowing the ferroelectric layer to be cleanly released while maintaining its crystalline quality and functional properties. Recent advances have expanded the repertoire of sacrificial layers to include Ca-doped SAO variants and super-tetragonal $Sr_4Al_2O_7$, which offer improved lattice matching capabilities and reduced crack formation during the release process[166]. These developments have enabled the fabrication of large-area, crack-free membranes spanning millimeter dimensions, a crucial requirement for practical device applications[35,169].

One of the most remarkable properties of freestanding ferroelectric membranes is their exceptional mechanical flexibility, which enables bending radii as small as a few micrometers without structural failure[170,171]. This extraordinary flexibility stems from the absence of substrate constraints and the inherent resilience of the perovskite crystal structure under mechanical deformation. For instance, $PbZrO_3$ membranes have demonstrated shape recoverability when bent to radii of curvature approaching 2.4 μm, corresponding to bending strains exceeding 2.5%[170]. This mechanical robustness, combined with preserved ferroelectric functionality, makes these membranes ideal candidates for flexible electronics and wearable devices[172,173].

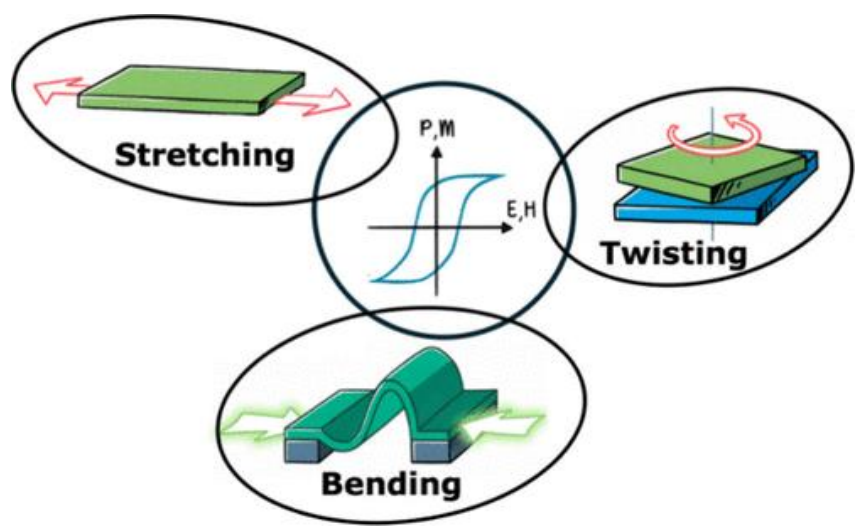

**Figure 7.** This schematic illustrates the diverse techniques employed for modulating the properties of freestanding oxide membranes and ferroic materials. These techniques encompass stretching, bending, and twisting. The central hysteresis loop illustrates the ferroic characteristics, such as polarization versus electric field (PE) or magnetization versus magnetic field (MH). Reproduced with permission.[174] Copyright 2025, American Chemical Society.

The mechanical flexibility of freestanding membranes introduces entirely new physics through flexoelectricity, the coupling between strain gradients and electric polarization[175–177]. When a ferroelectric membrane is bent, substantial strain gradients develop across its thickness, generating flexoelectric fields that can significantly modify the polarization landscape[176]. These strain gradients can reach magnitudes exceeding $10^6$ m$^{-1}$, orders of magnitude larger than those achievable in rigid films[175]. The resulting flexoelectric effects have been shown to stabilize novel domain configurations, enhance piezoelectric responses, and even induce ferroelectric-like behavior in nominally paraelectric materials such as $SrTiO_3$[178,176]. This flexoelectric coupling provides an additional degree of freedom for controlling ferroelectric properties and designing strain-programmed devices[179].

Strain engineering in freestanding membranes extends far beyond simple bending deformation. Researchers have demonstrated continuous uniaxial tensile strains exceeding 6% in $PbTiO_3$ membranes, values that are impossible to achieve in substrate-constrained films[180]. Under such large strains, dramatic structural phase transitions occur, including the rotation of polarization from out-of-plane to in-plane orientations and the emergence of complex domain structures with enhanced electromechanical coupling[180]. These strain-induced phenomena have revealed the existence of topological ferroelectric structures, including dipole spirals that exhibit giant piezoelectric responses exceeding 320 pC/N, opening new possibilities for ultra-sensitive sensors and high-performance actuators[167].

Recent experimental investigations have uncovered the emergence of exotic polarization textures in strained ferroelectric membranes. Phase-field simulations and experimental observations have revealed that bending-induced strain gradients can stabilize vortex-antivortex domain pairs, polar head-tail boundaries, and other topological structures that are energetically unfavorable in unstrained systems[171,179]. These findings have profound implications for understanding the fundamental limits of ferroelectric domain engineering and provide new pathways for creating materials with tailored functional properties.

The integration of freestanding ferroelectric membranes with two-dimensional materials has emerged as a particularly promising research direction[165]. Van der Waals heterostructures combining ferroelectric oxide membranes with semiconducting transition metal dichalcogenides have demonstrated exceptional performance in field-effect transistors, exhibiting mobilities comparable to hexagonal boron nitride-based devices while providing the additional functionality of ferroelectric switching[165]. These hybrid devices leverage the high dielectric constant of ferroelectric oxides to effectively screen Coulomb scattering centers while enabling non-volatile memory operation through polarization switching.

Applications of freestanding ferroelectric membranes span a broad spectrum of technologies, from flexible electronics to energy harvesting and sensing[38,172]. Piezoelectric nanogenerators based on freestanding $PbZr_{0.52}Ti_{0.48}O_3$ membranes have achieved record-high power densities exceeding 63 mW/cm$^3$, demonstrating exceptional energy conversion efficiency coupled with mechanical durability over tens of thousands of cycling operations[172]. These performances represent significant advances over conventional ceramic-based devices and highlight the potential for membrane-based technologies in wearable electronics and IoT applications. Furthermore, ferroelectric tunnel junctions based on freestanding $BaTiO_3$ barriers have shown promising results for flexible memory applications, maintaining robust tunneling electroresistance effects even under mechanical bending[181].

Beyond energy and memory applications, freestanding ferroelectric membranes have demonstrated unique capabilities in pressure sensing and mechanical actuation[182]. The enhanced strain sensitivity arising from flexoelectric coupling enables ultra-low pressure detection thresholds, with switching pressures as low as 0.06 GPa demonstrated in $PbTiO_3$ membranes[182]. This exceptional pressure sensitivity, combined with the non-volatile nature of ferroelectric switching, opens new possibilities for developing highly sensitive mechanical sensors and actuators.

Despite these remarkable advances, several challenges remain in the development of freestanding ferroelectric membranes. Controlling crack formation during the release process remains a critical issue, particularly for large-area applications[35,169]. While strategies such as capping layers and optimized sacrificial layer compositions have shown promise, achieving uniform, defect-free membranes across wafer-scale dimensions continues to require further development. Additionally, the long-term stability of these membranes under operational conditions, including thermal cycling and environmental exposure, requires a comprehensive investigation to ensure reliable device performance.

Future research directions in freestanding ferroelectric membranes are likely to focus on several key areas. The development of new sacrificial layer materials with improved lattice matching and water solubility could enable the fabrication of defect-free membranes for a broader range of ferroelectric compositions[183]. Advanced strain engineering techniques, including controlled wrinkling and origami-inspired folding, may unlock new topological states and functionalities[171]. The integration of ML approaches for optimizing membrane properties and predicting novel domain configurations represents another promising avenue for accelerating materials discovery and device optimization. Moreover, the exploration of ultrathin hafnia-based freestanding membranes, which have recently demonstrated ferroelectricity down to 1-nm thickness, may enable new classes of ultra-high-density memory devices[184].

In conclusion, freestanding ferroelectric membranes have established themselves as an important platform for exploring unconventional ferroelectric phenomena and developing next-generation flexible devices. Their unique combination of mechanical flexibility, enhanced electromechanical responses, and compatibility with diverse substrates positions them at the forefront of materials science research, with the potential to enable important technologies in electronics, energy, and sensing applications. The continued development of fabrication techniques, understanding of fundamental ferroelectric physics, and exploration of novel applications will likely cement their role as a cornerstone technology for the future of flexible and wearable electronics.

## 3. Multiferroics

Multiferroic materials, which exhibit the coexistence of two or more ferroic orders such as ferroelectricity, ferromagnetism, and ferroelasticity, have garnered significant attention due to their potential for device functionalities arising from the coupling between these orders. Recently, the integration of valley polarization and novel magnetic states, specifically magnetic skyrmions and altermagnetism, into multiferroic systems has opened up new avenues for manipulating magnetic properties with electric fields and vice versa. This section will delve into the recent progress in multiferroic materials, covering both intrinsic single-phase multiferroics and engineered composite multiferroics, such as heterostructures, incorporating the novel phenomena, and highlighting the fundamental physics, material realizations, and potential applications.

### 3.1. Novel magnetism (skyrmion and altermagnetism) and ferroelectricity

Then we focus on the interplay of ferroelectricity with magnetic skyrmions and altermagnetism, which is driven by the intense current research interest in achieving robust, energy-efficient electrical control over these novel, non-collinear, or compensated magnetic states. Such control is deemed crucial for advancing next-generation spintronic devices, and ferroelectricity provides a natural pathway for electric field manipulation.

*3.1.1. Skyrmion and ferroelectricity*

Magnetic skyrmions are topologically protected, swirling spin textures that exist at the nanoscale, shown in Figure 8[185]. They are characterized by a non-trivial topological charge, making them robust against perturbations and offering unique properties for device applications. The stability of skyrmions arises from the delicate balance between several competing magnetic interactions. The Dzyaloshinskii-Moriya interaction (DMI), an antisymmetric exchange interaction arising from spin-orbit coupling and broken inversion symmetry, plays a critical role in their formation. The interplay between DMI, magnetic anisotropy, and exchange interactions determines the size, shape, and stability of skyrmions. The small size and topological protection of skyrmions make them attractive for applications in high-density data storage and logic devices. Their ability to be manipulated by electric currents or magnetic fields further enhances their potential for applications[186]. The dynamics of skyrmions, particularly their current-driven motion, is also a key area of research, with implications for developing novel memory devices[187]. The exploration of skyrmions extends beyond fundamental physics, with significant technological implications for low-power spintronics.

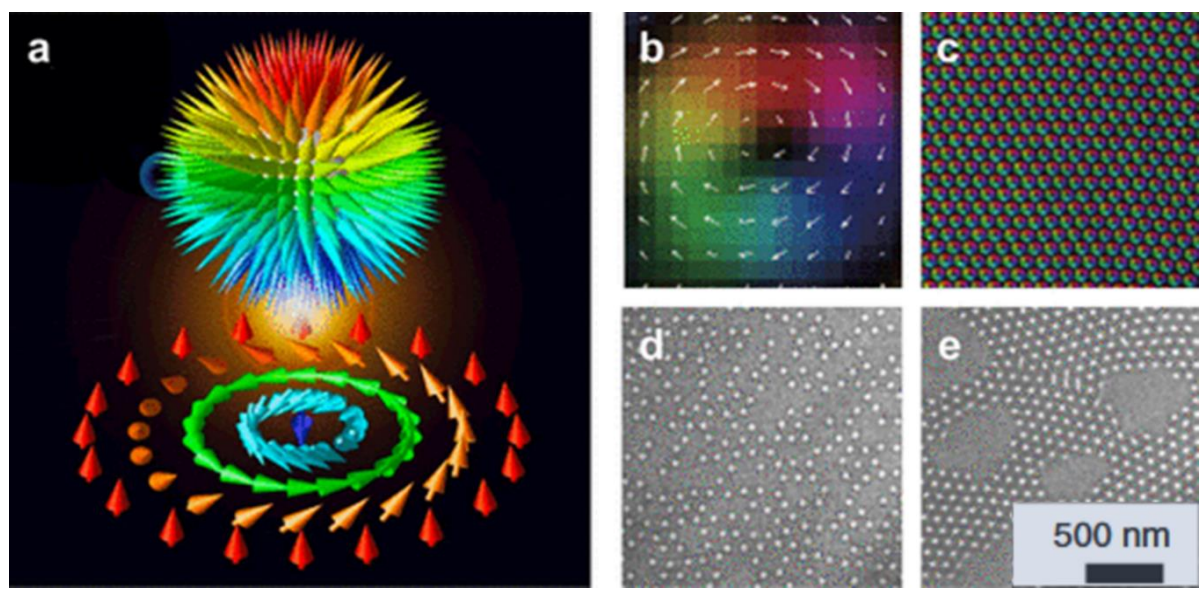

**Figure 8.** (a) Representation of topological charge (top) of the skyrmion (bottom). The constituent spin moments of the skyrmion wrap the sphere exactly one-time when the moment vectors are gathered with the common origin (upper panel). (b) In-plane magnetization configuration in a single skyrmion observed by Lorentz transmission electron microscopy. The colors represent the in-plane direction of the magnetization. (c–e) Various states of skyrmion condensates: crystal (c), isolated (d), and aggregated (e) states. Panels a are reproduced with permission from [185] Copyright 2020, American Chemical Society. Panels b–e are reproduced with permission from [188] Copyright 2018, Springer Nature Limited.

The combination of ferroelectricity and magnetism in multiferroic materials introduces the possibility of manipulating magnetic properties via electric fields, and vice versa, opening up avenues for novel device functionalities[189]. The ability to control the electric polarization non-volatile is essential for memory applications, and the integration of this control with magnetic skyrmions offers a path towards high-density, low-power spintronic devices[190,191]. Multiferroic heterostructures provide a powerful approach to achieving magnetoelectric coupling by combining ferromagnetic and ferroelectric materials[192–194]. This approach offers a high degree of control over the magnetic properties of the ferromagnetic layer through manipulation of the ferroelectric polarization in the adjacent layer. The interfacial interactions between the layers play a crucial role in determining the strength of the magnetoelectric coupling. The ability to create and annihilate skyrmions by switching the ferroelectric polarization has been demonstrated in several heterostructures[189,195–201]. This nonvolatile control of skyrmion states is highly desirable for memory applications. The design and optimization of multiferroic heterostructures require a careful consideration of the materials' properties, interface quality, and the underlying physical mechanisms responsible for the magnetoelectric coupling. The choice of materials is critical, as different combinations can lead to vastly different magnetoelectric coupling strengths and skyrmion properties. The exploration of novel heterostructures with enhanced magnetoelectric coupling and skyrmion stability remains a key area of ongoing research.

While multiferroic heterostructures are a common approach to achieving magnetoelectric coupling, some materials exhibit intrinsic multiferroicity, meaning that ferroelectricity and magnetism coexist within the same crystallographic phase. One notable example is $Cu_2OSeO_3$, which displays both skyrmion formation and ferroelectric properties[202,203,190,204]. However, only a few materials are known to host both robust ferroelectric order and intrinsic magnetic skyrmions concurrently, including $Cu_2OSeO_3$ and $GaV_4S_8$[205–207]. A key challenge in this area is the low Curie temperature often observed in these materials, restricting their practical applications[208]. The search for new single-phase multiferroics with higher transition temperatures and stronger magnetoelectric coupling is a critical area of research. The understanding of the microscopic mechanisms leading to intrinsic multiferroicity is crucial for the rational design of new materials with enhanced properties. This involves investigating the interplay of crystal structure, electronic structure, and spin-orbit coupling. The computational prediction and experimental verification of new single-phase multiferroics are vital steps towards realizing the full potential of this class of materials.

The Dzyaloshinskii-Moriya interaction (DMI), an antisymmetric exchange interaction, is paramount for the stabilization of chiral magnetic skyrmions, arising from spin-orbit coupling (SOC) in environments with broken inversion symmetry. In multiferroic systems, particularly heterostructures combining ferroelectric and magnetic layers, the interplay between ferroelectric polarization, strain, and applied electric fields can profoundly influence the DMI, offering pathways for electrical control of skyrmions. Ferroelectric polarization itself breaks inversion symmetry. When a ferroelectric material is interfaced with a magnetic layer, the polarization state of the ferroelectric can modify the local symmetry at the interface, thereby altering the strength and even the sign of the interfacial DMI. Theoretical models and first-principles calculations have shown that the magnitude of DMI can be linearly or non-linearly dependent on the polarization magnitude or specific ionic displacements. The strength of DMI, which can be theoretically obtained by four-state method[209,210], is highly sensitive to various factors, including the crystal structure, the presence of heavy elements, and interfacial effects in heterostructures[189]. In multiferroic systems, the ferroelectric polarization can influence the DMI, providing a pathway for electric-field control of skyrmion formation and stability[211]. Applied strain, whether biaxial or uniaxial, modifies the crystal lattice, altering bond lengths and angles[191]. These structural changes directly impact the magnetic exchange interactions, including the DMI. The ability to tune DMI through external stimuli such as electric fields or strain is essential for manipulating skyrmion properties in multiferroic devices. The relationship between DMI strength, magnetic anisotropy, and skyrmion stability has been extensively studied, providing valuable insights for designing materials and devices with optimized properties. Theoretical models and first-principles calculations are instrumental in understanding the microscopic origin of DMI and its dependence on various parameters[148] with tools like PASP (property analysis and simulation package for materials)[147]. An important theoretical framework for electric-field control of magnetic topological states was introduced by Xu et al., who proposed the EPDQ (Electric field-Polarization-DMI-topological charge) mechanism in type-I multiferroics[10]. This mechanism demonstrates that electric field switching of magnetic topological charge can be achieved through the mediation of electric polarization and DMI in a controllable and reversible fashion. EPDQ was validated by the multiferroic $VOI_2$ monolayer, which was predicted to host magnetic bimerons—topological objects consisting of two merons with opposite vorticity. Crucially, changes in magnetic anisotropy were identified, triggered by ferroelectric switching, play a fundamental role in realizing the EPDQ process, providing a novel pathway for addressing the long-standing challenge of electric-field control of magnetism. Recent studies have demonstrated practical implementations of the EPDQ-like mechanisms in van der Waals heterostructures, with Ferroelectric polarizations, engineered reversible skyrmion-bimeron switching achieved in $RuClBr/Ga_2S_3$ systems[212]. Similarly, comprehensive manipulation of magnetic topological textures has been realized in CrISe/In$_2$Se$_3$ heterostructures, where perpendicular strain and electric polarization enable field-free creation, annihilation, and conversion between skyrmions and bimerons[201] Experimental validation has also progressed, with hysteretic responses of skyrmion lattices to electric fields observed in magnetoelectric Cu$_2$OSeO$_3$, demonstrating nonvolatile and energy-efficient control of hexagonally packed skyrmion lattices[213]. These developments collectively establish electric-field control as a viable route toward practical skyrmion-based spintronic devices, building upon the foundational EPDQ mechanism.

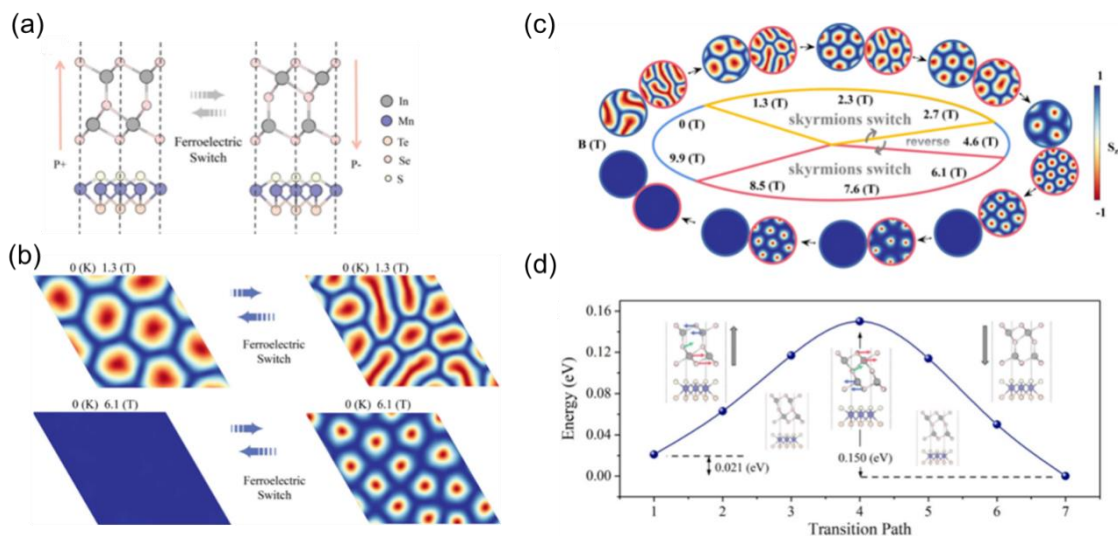

**Figure 9.** (a) Crystal structures of the MnSTe/In$_2$Se$_3$ heterobilayer under the $P^+$ and $P^-$ configurations. (b) Spin textures for the $P^+$ (left panel) and $P^-$ (right panel) phases under an external magnetic field of 1.3 T (upper panel) and 6.1 T (bottom panel) at 0 K. (c) Spin textures for the $P^+$ (with a blue edge) and $P^-$ (with a red edge) phases under various external magnetic fields at 0 K. (d) Ferroelectric transition pathway and energy barrier between the $P^+$ and $P^-$ phases of the heterobilayer. Reproduced with permission.[214] Copyright 2017, American Chemical Society.

The ability to control skyrmion states using electric fields is a key goal in multiferroic research[186,190,191,211,215], showing that switching the ferroelectric polarization can induce changes in DMI and magnetic anisotropy can lead to the creation, annihilation, or manipulation of skyrmions. This electric-field control offers a nonvolatile and energy-efficient way to manipulate skyrmions, which is highly desirable for memory applications. As shown in Figure 9, the ferroelectrics switching in MnSTe/In$_2$Se$_3$ heterobilayer introduce different spin textures, including skyrmions. The precise mechanism by which electric fields influence skyrmion states varies depending on the material system and the nature of the magnetoelectric coupling. This mechanism involves changes in the electronic structure, interfacial charge transfer, or strain-mediated effects. The ability to achieve reliable and repeatable electric-field switching of skyrmion states is a significant step towards the development of functional multiferroic devices[191]. The integration of electric-field control with other manipulation techniques, such as current-driven motion, offers even greater flexibility in controlling skyrmion dynamics[196,215–217]. The potential for high-density, low-power memory devices based on this approach is a major driver of research in this area. Further research is needed to optimize materials and device architectures for improved performance and reliability.

Mechanical strain provides another avenue for manipulating magnetic interactions and skyrmion properties in multiferroic systems[197,201,218]. Applying strain can alter the crystal lattice parameters, influencing the exchange interactions, DMI, and magnetic anisotropy. This strain-mediated control can lead to significant changes in skyrmion size, stability, and dynamics. Compressive strain, for example, has been shown to enhance DMI, leading to the formation of smaller and more stable skyrmions[197]. The precise control over strain magnitude and direction is essential for achieving desired skyrmion properties. This approach complements electric-field control, providing additional degrees of freedom for manipulating skyrmion dynamics. Further research is necessary to fully understand the interplay between strain, electric fields, and magnetic interactions in multiferroic systems and to optimize strain engineering techniques for device applications.

The unique properties of skyrmions in multiferroic materials hold immense potential for applications in spintronics[219]. Their small size and topological protection make them ideal candidates for high-density data storage. The ability to control skyrmions using electric fields offers a pathway to energy-efficient spintronic devices. Juge *et. al.* showcased the stabilization of antiferromagnetic skyrmions at room temperature within specially designed synthetic antiferromagnets (SAF) by carefully tuning the interlayer exchange coupling and chiral interactions[220]. Their research went further, demonstrating the ability to control these isolated skyrmions using electrical methods—a significant improvement over traditional ferromagnetic skyrmions, which often suffer from deflections due to dipolar fields. While SAFs themselves are not typically ferroelectric[221], their integration into multiferroic heterostructures with a ferroelectric layer offers a compelling route for electrical control of SAF-skyrmions. These findings highlight the exciting potential of SAF-based skyrmions for creating energy-efficient spintronic devices that can operate effectively at room temperature. The development of skyrmion-based racetrack memory devices, where skyrmions are moved along a magnetic track to represent data, is a major area of research. As demonstrated by Dohi *et. al.*[222], the efficient movement of skyrmions under low-current densities in thin-film multilayers enables high-density and reliable data storage. Skyrmions can also be used in logic devices, where their interactions can be exploited to perform logical operations. The robustness of skyrmions against thermal fluctuations is crucial for their application in devices operating at higher temperatures[216]. The exploration of novel skyrmion-based device architectures and the optimization of materials and fabrication techniques are essential steps towards realizing the full potential of skyrmions in spintronics. The development of reliable and scalable fabrication methods for skyrmion-based devices is a significant challenge.

Despite the significant progress in the field of multiferroics with skyrmions and ferroelectricity, several challenges remain. One major challenge is the identification and synthesis of new multiferroic materials with enhanced properties. This includes the search for materials with higher Curie temperatures, stronger magnetoelectric coupling, and improved skyrmion stability. The development of efficient and reliable methods for controlling skyrmion dynamics is also crucial. This requires a better understanding of the underlying physical mechanisms governing skyrmion behavior and the development of new manipulation techniques. The exploration of novel device architectures that leverage the unique properties of skyrmions in multiferroic systems is another important area of research[223]. The integration of multiferroic materials with existing semiconductor technologies is a key challenge for realizing practical applications. The development of scalable and cost-effective fabrication methods for multiferroic devices is also essential for commercial viability[196]. Further research is needed to overcome these challenges and to unlock the full potential of multiferroic materials with skyrmions and ferroelectricity for next-generation spintronic devices.

The unique properties of skyrmions, particularly when integrated into multiferroic systems, hold immense potential for applications in spintronics[224]. Their small size and topological protection make them ideal candidates for high-density data storage. The ability to control skyrmion creation, annihilation, and motion using electric fields via ferroelectric components in multiferroic structures, as discussed in this section, offers a pathway to energy-efficient spintronic devices.

*3.1.2. Altermagnetism and ferroelectricity*

The conventional understanding of magnetism encompasses two primary phases: ferromagnetism and antiferromagnetism[14]. Ferromagnets exhibit spontaneous magnetization, resulting in a net magnetic moment even in the absence of an external field. Antiferromagnets, on the other hand, possess zero net magnetization due to the antiparallel alignment of their spins. The emergence of altermagnetism challenges this established dichotomy[14]. Altermagnetism represents a distinct magnetic phase characterized by the coexistence of spin polarization and zero net magnetization, shown in Figure 10. This unique feature distinguishes it from both ferromagnets and antiferromagnets. The absence of net magnetization prevents the generation of stray fields, a significant advantage for device applications[41]. The origin of altermagnetism lies in the specific arrangement of spins dictated by the crystal's inherent symmetry. This often involves rotational operations that are not present in conventional ferromagnets or antiferromagnets[225].

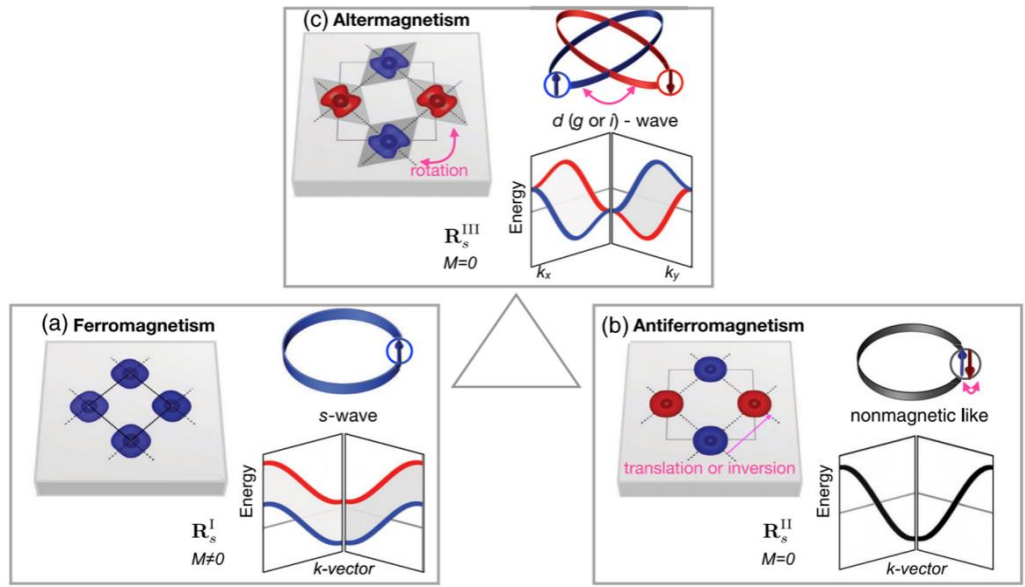

**Figure 10.** Illustrative models of collinear ferromagnetism, antiferromagnetism, and altermagnetism in crystal-structure real space and nonrelativistic electronic-structure momentum space. Reproduced with permission.[14] Copyright 2017, American Physical Society.

The nonrelativistic spin group formalism provides a fundamental framework for understanding the unique spin polarization characteristics of altermagnets[14]. This formalism incorporates both spin and spatial transformations, allowing for a more complete description of the magnetic order than traditional approaches. The crucial aspect of this formalism is the relationship of spin-polarized band structure between spin and momentum, expressed as $E_\uparrow(k) \neq E_\downarrow(k)$ but $E_\uparrow(k) = E_\downarrow(Rk)$, where $\uparrow$ and $\downarrow$ represent spin direction, $k$ represents momentum, and $R$ represents the spin-space rotation operation. This equation highlights how rotational symmetry operations fundamentally link the spin and momentum spaces in altermagnets.

The combination of altermagnetism and ferroelectricity leads to a new type of multiferroic material[41–43,225–229]. To understand this new multiferroics, it's essential to first examine the limitations of existing classifications[230–232,101]. Type-I multiferroics are characterized by ferroelectricity and magnetism having different origins and appearing largely independently, though with some coupling between them. Type-II multiferroics, in contrast, are materials where magnetism directly causes ferroelectricity, resulting in strong coupling between the two order parameters. Unlike Type-I and Type-II multiferroics, this new type of multiferroic exhibits an inherent coupling between altermagnetism and ferroelectricity mediated by crystal symmetry. This intrinsic coupling is the defining characteristic of the new type of multiferroics, making them particularly promising for applications requiring strong magnetoelectric coupling. Theoretical studies have shown that ferroelectric switching can effectively invert the spin polarization in altermagnets shown in Figure 11[43]. This direct coupling by symmetry operations between ferroelectric and magnetic orders is a key feature of this new type of multiferroics and is absent in the other types. The ability to fully control the magnetic order parameter using an electric field is a significant potential advantage for device applications. This contrasts sharply with traditional multiferroics, where the coupling is often weak and difficult to control.

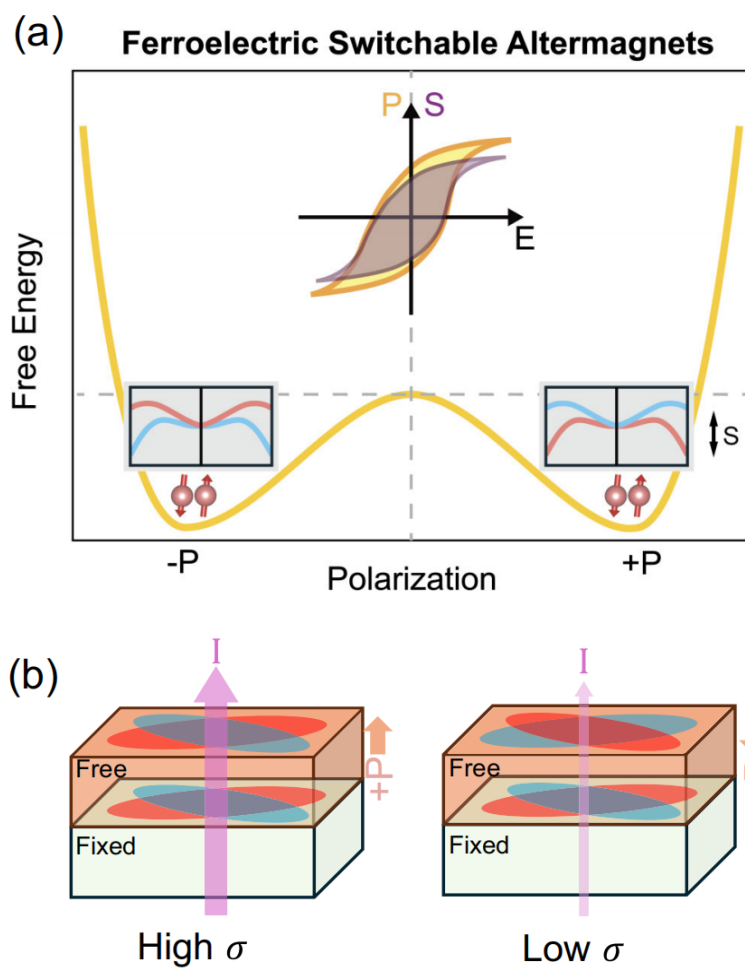

**Figure 11.** (a) Schematic representation of ferroelectrically switchable altermagnetism, in which the altermagnetic spin splitting S is strongly coupled to the ferroelectric polarization P.

(b) A design of a nonvolatile spin filtering tunnel junction device incorporating ferroelectrically switchable altermagnets. The Fermi surfaces of the fixed layer and the free layer are also depicted. The electric polarization P can be utilized to regulate the spin polarization of the free layer, thereby resulting in a transition between high- and low-conductance (σ) states. Reproduced with permission.[43] Copyright 2025, American Physical Society.

Bilayer $MnPSe_3$ has emerged as a leading example of the new type of multiferroic exhibiting robust magnetoelectric coupling[41]. Theoretical calculations and experimental findings strongly suggest that ferroelectric switching in this material leads to a complete inversion of the spin polarization. This observation directly demonstrates the strong, symmetry-driven magnetoelectric coupling predicted for this new type of multiferroic. The use of the magneto-optical Kerr effect (MOKE) provided crucial experimental verification of the predicted behavior[233,41]. The observed coupling strength in bilayer $MnPSe_3$ is significantly higher than that typically observed in traditional multiferroics, highlighting the potential of altermagnetic systems. The success of bilayer $MnPSe_3$ as a model system underscores the potential for designing other materials with similar properties and enhanced functionality. Further research into related materials and heterostructures is needed to fully explore the possibilities of this promising class of multiferroics.

$LiFe_2F_6$ represents another important material system demonstrating the coexistence of altermagnetism and ferroelectricity[234]. First-principles calculations predicted the multiferroic nature of $LiFe_2F_6$, identifying it as a $d$-wave altermagnet with charge-ordering-mediated ferroelectricity. The calculations revealed a direct correlation between the electronic structure, the magnetic order, and the electric polarization. Furthermore, the application of biaxial compressive strain induces a phase transition to a ferrimagnetic and ferroelectric phase, further emphasizing the strong magnetoelectric coupling in this material. The $d$-wave nature of the altermagnetism in $LiFe_2F_6$ is a key feature that distinguishes it from other altermagnets. The specific symmetry of the $d$-wave altermagnetism plays a crucial role in mediating the interaction with ferroelectricity. The material's unique electronic properties also allow for the emergence of spin-triplet excitonic insulator phases in the ferromagnetic state, adding another layer of complexity and potential functionality.

Beyond Bilayer $MnPSe_3$ and $LiFe_2F_6$, several other materials have shown promise in exhibiting the coexistence of altermagnetism and ferroelectricity. $BaCuF_4$ and $Ca_3Mn_2O_7$ are notable examples, exhibiting altermagnetic spin polarization with relatively high critical temperatures (275 K and 110 K, respectively)[225]. These higher critical temperatures are a significant advantage for practical applications, as they suggest the possibility of room-temperature operation. The observation of altermagnetism in these materials strengthens the case for the widespread occurrence of this phenomenon in ferroelectric systems. Symmetry analysis has also identified a substantial number of additional candidate materials for altermagnetic multiferroics[225]. This list includes $BiFeO_3$, a well-established multiferroic material, further highlighting the potential for finding a broader class of altermagnetic multiferroics. The identification of these candidates through symmetry analysis underscores the importance of theoretical predictions in guiding experimental efforts. The exploration of these candidate materials represents a crucial direction for future research in this field.

Meanwhile, the innovative concept of antiferroelectric altermagnets (AFEAM) is introduced and theoretically forecasted[235]. These materials exhibit a unique combination of antiferroelectricity and altermagnetism, leading to novel functionalities. First-principles calculations have demonstrated the feasibility of AFEAM in van der Waals metal thio(seleno)phosphates and perovskite oxides. The ability to manipulate the spin polarization in AFEAM using relatively weak electric fields is particularly promising for developing electrically controlled spintronic devices. It should be noted that AFEAM is a theoretical concept and has not been validated by experiments at this stage.

The integration of sliding ferroelectricity with altermagnetism in $SnS_2$/$MnPSe_3$/$SnS_2$ heterostructures provides a compelling example of a novel magnetoelectric coupling mechanism[226,233]. In these heterostructures, changes in lattice symmetry induced by the sliding ferroelectric transition trigger phase transitions between antiferromagnetism and altermagnetism in the $MnPSe_3$ layer[226]. This demonstrates a distinct magnetoelectric coupling mechanism mediated by lattice symmetry rather than solely by electronic effects. The use of heterostructures allows for the controlled manipulation of both ferroelectric and magnetic properties. The ability to trigger altermagnetism through changes in lattice symmetry opens new avenues for designing multiferroic devices. The precise control over the lattice symmetry and the resulting magnetic phase transitions offer significant advantages for creating advanced spintronic devices. This approach complements other methods for achieving magnetoelectric coupling, offering a diverse range of design strategies.

The ability to manipulate spin polarization using electric fields, a key feature of altermagnetic multiferroics, is highly desirable for developing new spintronic devices[225,229]. The altermagnetoelectric effect and the behavior of AFEAM demonstrate the feasibility of this electric-field control. This contrasts with traditional spintronic devices that rely on magnetic fields for control, often resulting in slower switching speeds and higher power consumption.

Electrically controlled spintronic devices based on altermagnetic multiferroics promise faster switching times and lower energy consumption. The potential for all-electric control of spintronic devices offers significant advantages for scalability and integration into existing electronic systems. This ability to manipulate spin states using electric fields could lead to faster, more efficient, and more versatile spintronic devices. The lack of stray fields in altermagnets is another significant advantage for high-density integration.

The strong magnetoelectric coupling inherent in altermagnetic multiferroics opens exciting possibilities for developing novel sensors and actuators[43,236]. These devices could leverage the interplay between electric and magnetic fields to detect or generate specific responses. For example, sensors could detect changes in magnetic fields by measuring the resulting changes in electric polarization, or actuators could generate precise movements by manipulating the magnetic order using electric fields. The strong coupling and the potential for all-electric control offer advantages in terms of sensitivity and efficiency. These applications could range from highly sensitive magnetic field sensors to energy-efficient actuators for various microelectromechanical systems. The versatility of altermagnetic multiferroics suggests a wide range of potential applications beyond spintronics, encompassing areas such as sensing, actuation, and other multifunctional devices.

The discovery of altermagnetism has significantly broadened the landscape of multiferroic materials. The unique combination of altermagnetism and ferroelectricity offers unprecedented opportunities for creating materials with strong magnetoelectric coupling, zero stray fields, and potentially high thermal stability. These properties are highly desirable for advanced spintronic and other device applications. While significant challenges remain in identifying new materials, understanding the underlying mechanisms, and developing efficient fabrication techniques, the potential of altermagnetic multiferroics for next-generation devices is substantial.

### 3.2. Ferrovalley and ferroelectricity

Ferrovalley materials represent a relatively new class of materials characterized by the presence of spontaneous valley polarization[16,45,46,237,47]. Valley polarization refers to the unequal population of electrons in different valleys of the material's Brillouin zone, shown in Figure 12. These valleys are distinct energy minima in the electronic band structure, often located at specific high-symmetry points (e.g., K points in hexagonal lattices). The existence of spontaneous valley polarization implies a broken time-reversal symmetry. Unlike spin or charge, valley polarization offers an additional degree of freedom for information storage and manipulation in electronic devices, leading to the emergence of valleytronics[237]. The control and manipulation of valley polarization are crucial for harnessing the potential of ferrovalley materials. Several methods have been proposed and demonstrated, including the application of electric fields, strain engineering, and magnetic fields. Electric fields can directly influence the band structure and shift the energy levels of different valleys, leading to changes in valley polarization. Strain engineering can modify the crystal symmetry and band structure, thereby influencing valley polarization. Magnetic fields, through the Zeeman effect, influence spins and, via spin-valley locking, induce valley polarization. The precise control of valley polarization opens up exciting possibilities for the design of novel valleytronic devices[237]. Further research into the underlying mechanisms of valley polarization and the development of efficient control methods is essential for advancing the field of valleytronics.

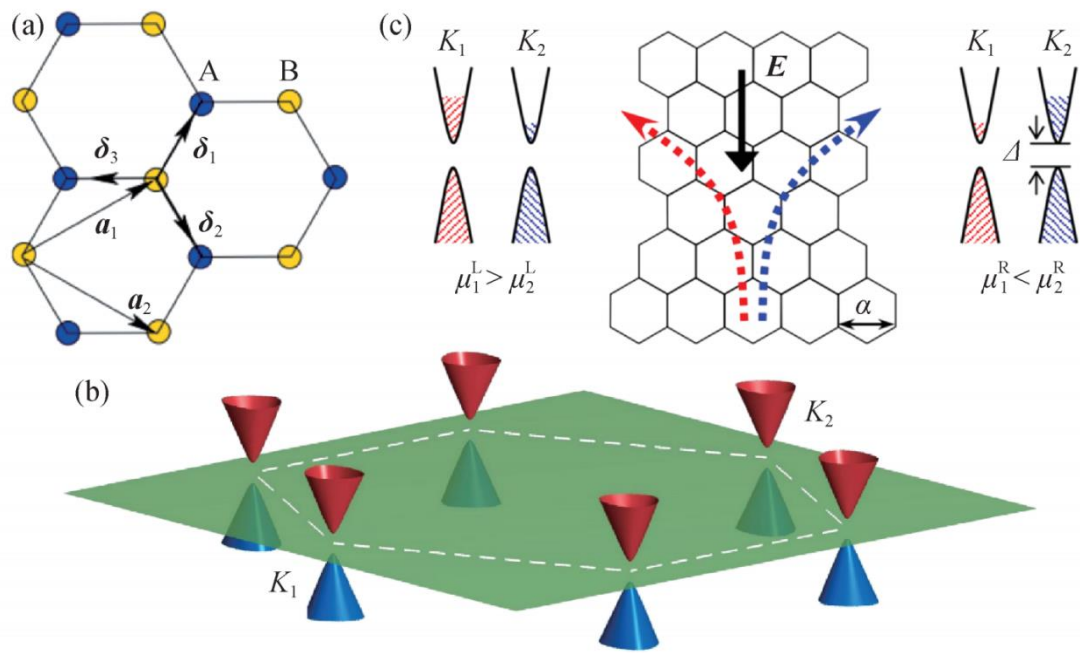

**Figure 12.** (a) In the honeycomb lattice of graphene, sublattices A and B are respectively represented by unfilled and filled circles. Vectors a1 and a2 are the lattice vectors. (b) Brillouin zone and high symmetry point. (c) In the upper panel is the band structure of graphene, while in the lower panel is the orbital magnetic moment of the conduction bands of graphene with broken spatial inversion symmetry. (d) Schematic diagram of the valley Hall effect. The in-plane electric field gives rise to a transverse valley current, which leads to a net valley polarization at the edge of the sample. Reproduced with permission.[16] Copyright 2023, China Research Institute of Radiowave Propagation.

The coexistence of ferroelectricity and ferrovalley polarization in a single material presents a particularly intriguing scenario[44–48,238,239]. The interplay between these two properties can lead to synergistic effects, where the ferroelectric polarization influences the valley polarization, and vice versa. The mechanisms underlying this interplay are complex and often involve intricate interactions between the crystal structure, electronic band structure, and spin-orbit coupling. For instance, the broken inversion symmetry associated with ferroelectricity can directly influence the Berry curvature, a key quantity determining valley polarization. Furthermore, external stimuli such as electric fields and strain can simultaneously affect both ferroelectric and ferrovalley properties, enabling control over both degrees of freedom. Several materials have been theoretically predicted or experimentally observed to exhibit this combined behavior[46]. For example, bilayer $YI_2$ with 3R-type stacking shows both valley polarization and ferroelectric polarization, demonstrating concurrent ferrovalley and multiferroic behaviors[45]. The understanding of the mechanisms and material examples of this synergistic interplay is still in its early stages. Further research, combining advanced theoretical modeling and experimental techniques, is needed to fully elucidate the underlying physics and explore the potential for new device applications.

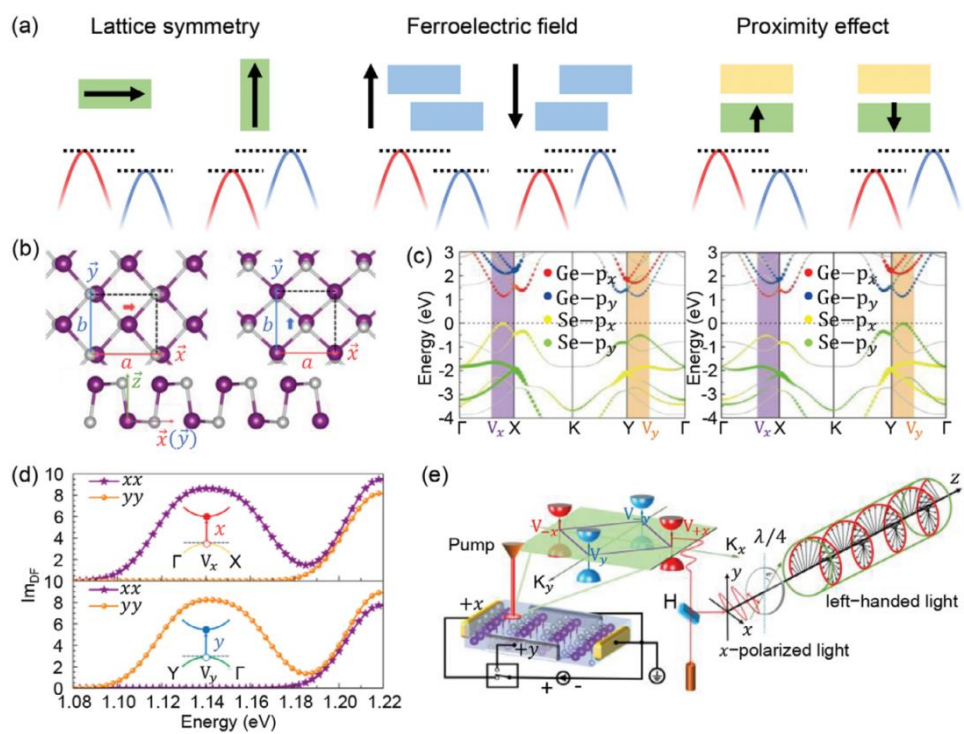

**Figure 13.** Ferroelectric-valley coupling in two-dimensional (2D) ferroelectric materials. (a) A schematic illustration depicting the lattice symmetry, ferroelectric field, and magnetic proximity effect within the context of ferroelectric-valley coupling. (b) The configurations. (c) The band structures. (d) The linear polarization absorption of the GeSe monolayer with ferroelectric polarization along the $x$-axis and $y$-axis. The term $Im_{DF}$ represents the imaginary

parts of dielectric functions. (e) A schematic diagram of the electrically tunable polarizer. Reproduced with permission.[240] Copyright 2017, IOP Publishing.

Electric fields provide a powerful means of controlling both ferroelectric and ferrovalley properties[44,240–242]. In ferroelectric materials, the application of an electric field can switch the direction of spontaneous polarization, leading to a change in the material's dielectric properties and other related effects. This switching behavior is often characterized by a hysteresis loop, which reflects the dependence of polarization on the applied electric field. In ferrovalley materials, electric fields can directly influence the band structure and shift the energy levels of different valleys, resulting in a change in valley polarization shown in Figure 13. The nonvolatile nature of ferroelectric switching makes it particularly attractive for applications in memory devices. In multiferroic materials exhibiting both ferroelectricity and ferrovalley, the electric field can provide a means for controlling both properties simultaneously, enabling the design of multifunctional devices. The electric field control of valley polarization is a relatively new area of research, and further investigation is needed to fully understand its potential for technological applications. The development of materials with enhanced electric-field controllability over valley polarization is a key goal for future research.

Strain engineering, the application of mechanical stress to modify the material's crystal structure and electronic properties, offers another powerful route for manipulating ferroelectric and ferrovalley properties[243–247]. The application of tensile or compressive strain can modify the crystal symmetry, leading to changes in the spontaneous polarization and the band structure. In ferroelectric materials, strain can induce phase transitions, alter the Curie temperature, and modify the piezoelectric response. In ferrovalley materials, strain can modify the valley degeneracy, shift the valley energies, and change the valley polarization. The degree of strain-induced modification depends on the material's elastic constants and the strength of the electron-phonon coupling. Strain engineering offers a particularly attractive method for tuning the properties of 2D multiferroic materials, as the van der Waals interactions between layers make them relatively susceptible to strain. Furthermore, strain can be applied locally, enabling the creation of complex patterns and functionalities. The ability to control both ferroelectric and ferrovalley properties through strain engineering opens exciting possibilities for developing new multifunctional devices.

Magnetic fields can significantly influence the ferrovalley properties of multiferroic material[248,244,45,46,238]. In materials where spin-valley locking exists, the application of a magnetic field can directly influence the valley polarization by aligning the spins and affecting the band structure, such as $ScBr_2$[239]. DMI can also play a crucial role in influencing valley polarization in magnetic multiferroics. The strength of the magnetic field effect depends on the material's magnetic anisotropy and the strength of the SOC. The control of valley polarization via magnetic fields offers a complementary approach to electric field control, providing additional flexibility in designing multifunctional devices. The interplay between magnetic order and valley polarization is a complex and fascinating area of research, with significant implications for spintronics and valleytronics. Further investigation into the mechanisms underlying this interplay is crucial for developing new materials and devices.

A promising approach for independently reading out a ferrovalley state, without interference from other secondary effects, valley-selective transport measurements, particularly by exploiting the anomalous valley Hall effect (AVHE)[249]. In systems such as p-type $VSe_2$, under moderate hole doping where the Fermi level lies between the valence band maxima of the inequivalent $K^+$ and $K^-$ valleys, the Berry curvature is sharply localized near the valleys and nearly vanishes around the Γ point. This enables valley-polarized carriers to contribute directly to transverse charge currents under an in-plane electric field, while carriers from the Γ region propagate longitudinally without deflection.

For instance, when a ferrovalley state favors the $K^+$ valley, the dominant spin-down holes acquire a transverse velocity due to the positive Berry curvature and accumulate on one side of the sample, generating a measurable transverse voltage. Upon reversal of the valley polarization (e.g., favoring the $K^-$ valley), spin-up holes become dominant and deflect oppositely, producing a charge Hall voltage with a reversed sign.

Specifically, a multifunctional memory device can be proposed that leverages the coexistence of ferroelectric and ferrovalley orders[47]. In this design, the ferroelectric polarization can be electrically switched, which in turn reverses the valley polarization due to the strong coupling between the two order parameters. The ferrovalley state can then be optically read out using circularly polarized light, as the two ferrovalley states couple selectively to left- and right-handed circular polarization. This electric-write, optical-read scheme offers a practical route for non-volatile memory applications combining ferroelectricity and valleytronics.

The unique combination of ferroelectricity and ferrovalley polarization in multiferroic materials opens up exciting possibilities for applications in valleytronics and spintronics. Valleytronics, which exploits the valley degree of freedom for information processing and storage, promises faster and more energy-efficient devices compared to conventional electronics. In multiferroic materials, the electric field control of ferroelectricity can be used to manipulate valley polarization, leading to the development of nonvolatile valleytronic devices. Spintronics, which utilizes the spin degree of freedom, offers similar advantages in terms of speed and energy efficiency. The interplay between ferroelectricity and magnetism in multiferroics allows for the electric field control of magnetic properties, leading to new possibilities for spin-based logic circuits and other advanced electronic systems. The combination of valleytronics and spintronics in multiferroic materials could lead to the development of highly advanced and multifunctional devices with enhanced performance. The integration of ferrovalley and ferroelectric properties in a single material opens up new avenues for exploration in this field.

## 4. Summary and Perspectives

The landscape of ferroic materials has evolved dramatically in recent years, propelled by discoveries that have expanded the boundaries of our understanding and opened new pathways for technological innovation. This review has delved into the multifaceted world of novel ferroic systems. From unconventional ferroelectrics to multiferroics integrating novel magnetic states and ferrovalley phenomena, the field has witnessed a possibility that redefines nanoelectronics, spintronics, and beyond.

The exploration of novel ferroic materials has yielded a diverse portfolio of systems with important potential. Hf-based ferroelectrics, notably $HfO_2$ and its doped variants, stand out for their robust ferroelectricity at nanoscale dimensions and seamless integration into CMOS processes, achieved through doping, oxygen vacancy engineering, and strain manipulation. Elementary ferroelectrics like Bi monolayers and Te nanowires showcase simplicity and scalability, leveraging lone-pair-driven distortions for room-temperature ferroelectricity. Stacking ferroelectrics, exemplified by bilayer h-BN and TMDs, offers tunable polarization via interlayer stacking, expanding the toolkit of 2D materials. Polar metal, including 2D $WTe_2$, merges conductivity with switchable polarization, challenging traditional paradigms. FQFE, such as α-$In_2Se_3$ and AgBr, introduce unconventional polarization mechanisms with low switching barriers, hinting at energy-efficient applications. Wurtzite-type ferroelectrics have exhibited robust ferroelectric switching and could play a pivotal role in piezoelectric and optoelectronic applications. Freestanding ferroelectric membranes—ultrathin sheets that maintain polar order without a substrate—demonstrate stable and switchable polarization even under extreme mechanical flexibility. Multiferroic systems integrating magnetic skyrmions, as in $Cu_2OSeO_3$ and $Fe_3GeTe_2$/$In_2Se_3$ heterostructures, enable electric-field control of magnetic textures, while ferrovalley-ferroelectric coupling in 2D materials harnesses valley degrees of freedom for electronic control. This rich material diversity, spanning oxides, elemental systems, and van der Waals heterostructures, underscores the field's breadth and its capacity to address varied technological needs.

The potential applications of these advancements are vast and poised to impact multiple domains. Hf-based ferroelectrics are set to advance non-volatile memory with Ferroelectric RAM (FeRAM), offering superior speed, endurance, and energy efficiency over flash memory, alongside neuromorphic computing through analog switching properties. Elementary ferroelectrics promise high-density data storage and self-gated transistors, capitalizing on their ultrathin nature and resistive switching capabilities. Stacking ferroelectrics provides a platform for flexible, ultrathin devices with tunable functionalities, ideal for next-generation nanoelectronics. Polar metals could enable fast-switching, low-power memory by integrating polarization with conductivity. The intrinsic ferroelectricity in wurtzite-type materials expands the toolkit for designing polarization-based devices and could enable integrated photonic and optoelectronic functionalities. Freestanding ferroelectric membranes provide opportunities for dynamic mechanical and electronic tuning. Multiferroics with magnetic skyrmions and altermagnetism offer pathways to ultra-low-power magnetic memory and logic devices, such as racetrack memory and spintronic gates, while ferrovalley-ferroelectric systems pave the way for valleytronics and quantum information processing. These applications, rooted in the unique properties of novel ferroics, highlight their potential to overcome limitations in current technologies and drive innovation in electronics and beyond.

Despite these strides, significant challenges persist that must be tackled to fully harness the promise of novel ferroic materials. Room-temperature functionality remains a significant obstacle for many of these materials. Addressing this will likely involve high-throughput computational and experimental screening of large chemical spaces (e.g., via ML models like SpinGNN[250,251] and HamGNN[252]), as well as advanced fabrication techniques to stabilize promising phases. Stabilizing ferroic phases remains a hurdle, particularly in unconventional systems. For Hf-based ferroelectrics, precise control over doping, oxygen vacancies, and strain is essential for consistent nanoscale performance, a task complicated by manufacturing variability. Elementary ferroelectrics face stability and scalability issues, requiring robust synthesis methods for integration into semiconductor platforms. Stacking ferroelectricity depends on exact interlayer configurations, sensitive to external perturbations, and demanding advanced fabrication precision. Polar metals struggle with fully switchable polarization in bulk forms due to screening effects, necessitating further breakthroughs in material design. FQFE requires deeper insight into its complex switching dynamics driven by large atomic displacements. Wurtzite ferroelectrics, while offering exceptional thermal stability and wide bandgaps that ensure low leakage currents, require strategic chemical doping to achieve practically manageable coercive fields, with optimization of composition and processing conditions remaining critical for CMOS integration. Freestanding ferroelectric membranes present unique opportunities with their dramatically enhanced piezoelectric responses exceeding bulk materials, yet face fabrication challenges in achieving reliable substrate release and maintaining structural integrity while preserving their enhanced electromechanical properties. In multiferroics, enhancing the coupling strength between ferroic orders, especially at room temperature, is critical for practical devices, as is stabilizing magnetic skyrmions and other magnetic textures under operational conditions. These challenges underscore the need for continued innovation in synthesis, characterization, and theoretical modeling.

Looking forward, the field stands at a pivotal juncture with numerous opportunities on the horizon. High-throughput computational screening, powered by ML and first-principles methods[253–256,250,251,97], promises to accelerate the discovery of new ferroic materials by identifying candidates with tailored properties. The development of universal ML Hamiltonians[257,258] and potentials[259,260] will enable simulations of complex systems at unprecedented scales, enhancing predictive capabilities. AI-driven autonomous materials discovery platforms, which integrate high-throughput DFT, advanced ML potentials and Hamiltonians, and robotic synthesis/characterization, will dramatically accelerate the identification and optimization of novel ferroics with on-demand properties. This includes a push towards lead-free piezoelectrics with performance rivaling PZT, and room-temperature multiferroics exhibiting strong, deterministic magnetoelectric coupling. Furthermore, exploring ferroic phenomena in extreme environments or at ultrafast timescales may unveil new physics and functionalities. Focusing on these targeted efforts will be crucial for advancing the field beyond fundamental discoveries towards tangible applications in nanoelectronics, spintronics, and quantum technologies. As the boundaries of ferroic science continue to expand, the promise of advancing nanoelectronics, spintronics, and quantum technologies draws ever closer, suggesting that these materials may play an increasingly significant role in future technological advancements, provided that key research and development challenges are successfully addressed.

**Acknowledgments**

We acknowledge financial support from the National Key R&D Program of China (No. 2022YFA1402901), NSFC (grants No. 12188101), Shanghai Science and Technology Program